\documentclass[12pt]{article}
\usepackage{amsfonts}
\usepackage{latexsym}
\usepackage{amsmath}
\usepackage{amssymb}
\usepackage{amssymb}

\allowdisplaybreaks[1]

\newcommand{\newsection}{    
\setcounter{equation}{0}\section}
\def\appendix#1{\addtocounter{section}{1}\setcounter{equation}{0}
\renewcommand{\thesection}{\Alph{section}}
\section*{Appendix \thesection\protect\indent \parbox[t]{11.15cm}{#1}}
\addcontentsline{toc}{section}{Appendix \thesection\ \ \ #1}}

\newcommand{\be}{\begin{eqnarray}}
\newcommand{\ee}{\end{eqnarray}}
\newcommand{\bea}{\begin{eqnarray}}
\newcommand{\eea}{\end{eqnarray}}
\newcommand{\ba}{\begin{array}}
\newcommand{\ea}{\end{array}}
\newcommand{\nn}{\nonumber \\}

\newenvironment{frcseries}{\fontfamily{frc}\selectfont}{}

\def \la {\label}

\def\a{\alpha}

\def\l{\lambda}

\def\e{\epsilon}

\def\hn{{\hat{\nu}}}

\def\bbe{{\bf{e}}}
\font\mybb=msbm10 at 11pt

\def\bb#1{\hbox{\mybb#1}}

\def\bR {\bb{R}}

\def\tn {{\tilde{\nabla}}}

\def\hn {{\hat{\nabla}}}

\begin{document}
\begin{titlepage}
\begin{center}
\vspace*{-1.0cm}
\hfill DMUS--MP--13/16 \\

\vspace{2.0cm} {\Large \bf Index theory and dynamical symmetry enhancement near IIB horizons} \\[.2cm]

\vspace{1.5cm}
 {\large  U. Gran$^1$, J. Gutowski$^2$ and  G. Papadopoulos$^3$}

\vspace{0.5cm}

${}^1$ Fundamental Physics\\
Chalmers University of Technology\\
SE-412 96 G\"oteborg, Sweden\\

\vspace{0.5cm}
$^2$ Department of Mathematics \\
University of Surrey \\
Guildford, GU2 7XH, UK \\

\vspace{0.5cm}
${}^3$ Department of Mathematics\\
King's College London\\
Strand\\
London WC2R 2LS, UK\\

\vspace{0.5cm}

\end{center}

\vskip 1.5 cm
\begin{abstract}

\end{abstract}
We show that the number of supersymmetries of IIB black hole horizons is $N=2 N_-+2\,\, \mathrm{index}(D_{\lambda})$, where $\mathrm{index}(D_\lambda)$ is the index of the  Dirac operator
  twisted with the line bundle $\lambda^{1\over2}$ of IIB scalars, and $N_-$ is the dimension of the kernel of a horizon Dirac operator which depends on IIB fluxes.  Therefore, all IIB horizons preserve an even number of supersymmetries. In addition if  the horizons have non-trivial fluxes and $N_-\not=0$, then $\mathrm{index}(D_\lambda)\geq 0$
and the horizons  admit an $\mathfrak{sl}(2,\bR)$ symmetry subalgebra. This provides evidence that all such horizons have an AdS/CFT dual.  Furthermore if the orbits of  $\mathfrak{sl}(2,\bR)$ are two-dimensional,
the IIB horizons are warped products  $AdS_2\times_w {\cal S}$.

\end{titlepage}



\setcounter{section}{0}
\setcounter{subsection}{0}
\setcounter{equation}{0}

\newsection{Introduction}

It has been known for some time that  black holes and  branes exhibit symmetry enhancement near their horizons \cite{carter, gibbons1, gwgpkt}. Typically,
this  enhancement leads to a superconformal symmetry  which in turn has been instrumental
in  the AdS/CFT correspondence \cite{maldacena} and black hole entropy counting \cite{strominger, sen}. Originally, this symmetry enhancement has been observed on an either a case by
case basis or under some additional symmetry assumptions, see \cite{lk} for a recent review and references within, but more recently it has been shown that it is a general phenomenon.
So far a proof has been given for odd-dimensional black hole horizons\footnote{ The black hole horizons in this paper
are assumed to be  Killing horizons with compact horizon sections. Such horizons can be event horizons for black holes, and brane configurations which after intersections and wrappings behave effectively as
0-branes. This justifies the requirement that the horizon section ${\cal S}$ is compact without boundary because ${\cal S}$ is thought of as the space that surrounds the point-like object. The associated black hole and brane solutions
 are not necessarily taken to be asymptotically flat.} like those of minimal 5-dimensional gauged supergravity and
of M-theory \cite{5index, 11index}.  It relies on the smoothness of horizons,
the compactness of the horizon sections,  Lichnerowicz type  theorems, and the vanishing of the index of the Dirac operator on odd dimensional manifolds. In particular, one shows first that there
is an enhancement of supersymmetry and then that the horizons with non-trivial fluxes  admit an $\mathfrak{sl}(2,\bR)$ symmetry subalgebra. In turn, this can be seen as
evidence that  all such horizons  have an AdS/CFT dual.

The investigation of the  symmetry enhancement of IIB horizons is expected to be different from that of M-horizons for two different reasons. First,  the index of the Dirac operator may not vanish
on even-dimensional  manifolds.
Second, IIB and the majority of other supergravity theories apart from the gravitino Killing spinor equation  (KSE) also have additional  KSEs associated with the rest of fermions in the spectrum.
Because of this, the Lichnerowicz type theorems needed to show symmetry enhancement must be generalized, so that all the Killing spinors, and not only the
solutions of the gravitino KSE, are in a one to one correspondence with the zero modes of a Dirac operator.

In this paper, we shall show that the number of Killing spinors of IIB horizons is given by
\bea
N=2 N_-+ 2\,\, \mathrm{index}(D_\l)~,
\la{index}
\eea
where $N_-$ is the dimension of the kernel of a  Dirac operator ${\cal D}^{(-)}$ on the horizon section ${\cal S}$ which depends on the IIB form fluxes, and  $D_\l$ is the twisted Dirac operator on ${\cal S}$
with respect to the $\lambda^{1\over2}$ line bundle  which arises in the description of IIB scalars.   In particular,
\bea
\mathrm{index}(D_\lambda)= \hat A\,\mathrm{ch}(\lambda^{1\over2})[{\cal S}]= {1\over 5760} (-4 p_2+7 p_1^2)-{1\over192} p_1 c_1^2+{1\over 384} c_1^4~,
\eea
where $\hat A$ is a A-roof genus, $\mathrm{ch}(\lambda^{1\over2})$ is the Chern character of $\lambda^{1\over2}$,  $p_1$ and $p_2$ are the first and second Pontryagin classes of ${\cal S}$, and $c_1$ is the first Chern class of $\lambda$. In many examples\footnote{If the IIB scalars take values in the hyperbolic upper half plane, then $c_1=0$ as it is a contractible space. See section \ref{ch5}
for the case where the scalars take values in the fundamental domain of the modular group.}
 $c_1=0$ and  the above formula gives the index of the
standard Dirac operator for 8-dimensional spin manifolds.

An immediate consequence of the formula (\ref{index}) is that the number of supersymmetries preserved by IIB horizons is even. Moreover, there are two  cases to consider depending on whether $N_-$ vanishes or not.  If $N_-\not=0$,  one  can show that $\mathrm{index}(D_\l)\geq 0$ and that the near horizon geometries with non-trivial fluxes admit an $\mathfrak{sl}(2, \bR)$ symmetry subalgebra.   In addition, the vector fields which generate the $\mathfrak{sl}(2, \bR)$ symmetry have either 2- or 3-dimensional orbits. If the orbits are 2-dimensional, then the near horizon
geometry is a warped product of $AdS_2$ with the horizon section ${\cal S}$, $AdS_2\times_w {\cal S}$.  This extends the results we have obtained
on symmetry enhancement for 5-dimensional and M-horizons to IIB horizons.

On the other hand if $N_-=0$, then the number of supersymmetries preserved is expressed in terms of the index
of the Dirac operator.  The geometry of
such horizons has already been investigated in \cite{iibhor}. The formula for $N$ in this case resembles that for the number of parallel spinors on 8-dimensional manifolds $M$ with (torsion-free)  holonomy strictly $Spin(7)$, $SU(4)$,  $Sp(2)$ and $Sp(1)\times Sp(1)$. In particular, it is known that the number of parallel spinors $N_p$ is given in terms of the
index \cite{atiyah1}  of the Dirac operator\footnote{On 8-dimensional manifolds, the vanishing of the topological obstruction $8 e-4 p_2+ p_1^2=0$ for the existence of a $Spin(7)$ structure and the expression for the signature $45 \sigma=7 p_2-p_1^2$  allows one to write
$\mathrm{index}(D)=\hat A[M]={1\over 5760} (-4 p_2+7 p_1^2)={1\over24}(-1+b_1-b_2+b_3+b^+_4-2b^-_4)$ \cite{joyce}, where $b_i$ are the Betti numbers,  $b^\pm_4$ are the number of self-dual and anti-self-dual Harmonic 4-forms, $\sigma=b_4^+-b_4^-$ and $e=2-2b_1+2b_2-2b_3+b_4$
is the Euler number.} as
\bea
N_p=\mathrm{index}(D)= \hat A[M]~,
\la{partop}
\eea
for $N_p=1,2, 3$ and $4$, respectively. Therefore $N_p$ is a topological invariant. In turn, (\ref{partop}) can be used to test whether a manifold with known first and second Pontryagin numbers
can admit a metric for which the Levi-Civita connection has holonomy one of the above four groups.

The formula (\ref{index}) also provides a topological restriction on the horizon sections. We know that $N\leq 32$.  As a result, one can conclude that $\mathrm{index}(D_\l)\leq 16-N_-$
with $0\leq N_-\leq 16$.  So for all horizon sections  $0\leq \mathrm{index}(D_\l)\leq 16$.  Furthermore since there are no IIB backgrounds with strictly $29, 30, 31$ supersymmetries
\cite{iib31} and the unique  solution \cite{iib28}  with $28$ supersymmetries \cite{bena} does not satisfy the
compactness restrictions of near horizon geometries, one also has that $ N_-+ \,\, \mathrm{index}(D_\l)\not=14, 15$.

It is well known that IIB supergravity consistently truncates to the common sector and to gravity coupled only to  the 5-form flux. In both cases $c_1=0$. We show that    common
sector horizons always preserve an even number of supersymmetries and that the index of the Dirac operator vanishes on the horizon sections.  Moreover, the geometry
of all common sector horizons can be understood in terms of that of heterotic horizons \cite{hhor}. In particular, for all common sector horizons, the orbit of the $\mathfrak{sl}(2, \bR)$ symmetry
is always 3-dimensional -- there are no common sector horizons which are warped $AdS_2$ products.

Turning to horizons with only 5-form flux, there are two cases to consider depending on whether $N_-$ vanishes. If $N_-\not=0$, and an additional  mild assumption when the orbit of
$\mathfrak{sl}(2, \bR)$ is 3-dimensional, all such horizons
preserve 4k supersymmetries and the index of the Dirac operator vanishes on the horizon sections. As a consequence the $\hat A$-genus of horizon sections vanishes. On the other hand
 if $N_-=0$, the number of supersymmetries preserved is a given in terms of the index of a Dirac operator and so it is a topological invariant. The geometry of such horizons has been investigated in
 \cite{iibfive}.

The proof of our results relies on first integrating the KSEs of IIB supergravity along the lightcone directions after decomposing the Killing spinor as $\e=\e_++\e_-$ using the lightcone projectors $\Gamma_\pm \e_\pm=0$,  and then showing that the independent KSEs
are those that are obtained by the naive restriction of IIB KSEs on the horizon section ${\cal S}$. The integration along the lightcone directions\footnote{To integrate the KSEs we do {\it not} use the bilinear matching condition which has been extensively applied to investigate near horizon geometries following \cite{reall} but which imposes an additional
restriction on the solutions, see also \cite{d4hor}.} gives rise
to two horizon supercovariant connections $\nabla^{(\pm)}$ on the  sections ${\cal S}$, and two associated horizon Dirac operators ${\cal D}^{(\pm)}$.  Furthermore, it establishes
a map $\Gamma_+\Theta_-$,  which depends on IIB fluxes, from the $\epsilon_-$ Killing spinors to the $\epsilon_+$ Killing spinors. The proof proceeds with the demonstration of two Lichnerowicz type theorems,
one for each  ${\cal D}^{(\pm)}$ horizon Dirac operator. These are established with the use of the (strong) maximum principle  for the ${\cal D}^{(+)}$ operator and with a partial
integration argument for ${\cal D}^{(-)}$. In both cases, the proof requires the field equations and Bianchi identities of IIB supergravity. The novelty in the proof of these two
Lichnerowicz type theorems is that in addition to implying  that the zero modes of ${\cal D}^{(\pm)}$ are $\nabla^{(\pm)}$-parallel, they also establish that these zero modes
also solve the two algebraic horizon KSEs  associated with the IIB dilatino KSE.
Then the formula (\ref{index}) is shown
as an application of the index theorem for the Dirac operator after observing that the number of Killing spinors is $N=N_++N_-$, where $N_\pm= \mathrm{dim}\, \mathrm {Ker}\, {\cal D}^{(\pm)}$, and demonstrating that the kernel of adjoint of  ${\cal D}^{(+)}$ can be effectively identified with that of  ${\cal D}^{(-)}$.

The proof of $\mathfrak{sl}(2,\bR)$ symmetry for horizons with $N_-\not=0$ utilizes the linear map $\Gamma_+\Theta_-$ mentioned above. In particular we show that for horizons
with non-trivial fluxes, $\Gamma_+\Theta_-$ is an injection and so for each zero mode of ${\cal D}^{(-)}$ there is a zero mode of ${\cal D}^{(+)}$. Using such a pair
of zero modes, one can construct three 1-form Killing spinor bi-linears and demonstrate that satisfy an $\mathfrak{sl}(2,\bR)$ algebra.

This paper has been organized as follows. In section 2, we integrate the KSEs along the lightcone directions and find the independent KSEs. In section 3, we outline the
proof of two Lichnerowicz type theorems for the horizon Dirac operators. In section 4, we prove (\ref{index}). In section 5, we relate the zero modes of the two
horizon Dirac operators and in section 6 we use this result to demonstrate the $\mathfrak{sl}(2, \bR)$ symmetry of some IIB horizons. In section 7, we
investigate the common sector and IIB horizons with only 5-form fluxes, and in section 8 we give our conclusions and state two conjectures. In appendix A, we summarize
the field equations and Bianchi identities we use through this paper, and in appendices B and C we provide detailed proofs of the theorems we have used. In appendix D a proof of the fact that the Killing vectors constructed from Killing spinors preserve the fluxes up to a $U(1)$ transformation is given.

\newsection{Horizon fields and KSEs}

\subsection{Horizon fields, Bianchi identities and field equations}

The fields of IIB supergravity near an extreme horizon can be expressed \cite{iibhor}  as
\be
ds^2 &=&2 \bbe^+ \bbe^- + \delta_{ij} \bbe^i \bbe^j~,~~~F= r \bbe^+ \wedge X + \bbe^+ \wedge \bbe^- \wedge Y + \star_8 Y~,~~
\cr
G &=& r \bbe^+ \wedge L + \bbe^+ \wedge \bbe^- \wedge \Phi + H~,~~~P = \xi~,
\la{hormetr}
\ee
where we have introduced the frame
\be
\label{basis1}
\bbe^+ = du, \qquad \bbe^- = dr + rh -{1 \over 2} r^2 \Delta du, \qquad \bbe^i = e^i_I dy^I~,
\ee
 and the self-duality of $F$ requires that  $X = - \star_8 X$.
The dependence on the coordinates $u$ and $r$ is explicitly given, and the horizon section ${\cal S}$ is the co-dimension 2 submanifold $r=u=0$. $\Delta$, $h$, $Y$ are 0-, 1- and 3-forms
on ${\cal S}$, respectively, $\Phi$, $L$ and $H$ are $\lambda$-twisted 1-, 2- and 3-forms on ${\cal S}$, respectively, and $\xi$ is a $\lambda^2$-twisted 1-form on ${\cal S}$, where $\lambda$ arises
 from the pull back of the canonical bundle on the scalar manifold $SU(1,1)/U(1)$ on ${\cal S}$. Throughout, we require that all
 field strengths are taken to be at least $C^2$ differentiable on ${\cal S}$ . This is one of our key assumptions. Furthermore, we shall not dwell on whether
 we consider $SU(1,1)/U(1)$ or the fundamental domain of the modular group as the IIB scalars manifold, however see section 5.

Substituting the fields (\ref{hormetr}) into the Bianchi identities and field equations of IIB supergravity, one finds that $L$ and $X$ can be expressed in terms
of other fields. The remaining Bianchi identities and field equations are summarized in appendix A.

\subsection{Horizon KSEs and lightcone integrability} \label{lightcone}

The gravitino and dilatino KSEs of IIB supergravity \cite{schwarz1, schwarz2}  are
\bea
\label{gkse}
&&\bigg(\nabla_M -{i \over 2} Q_M
+{i \over 48} F_{M N_1 N_2 N_3 N_4}\Gamma^{N_1 N_2 N_3 N_4} \bigg) \epsilon
\cr
&&~~~~~~~~~~~~-{1 \over 96} \bigg(\Gamma_M{}^{N_1 N_2 N_3} G_{N_1 N_2 N_3} -9 G_{M N_1 N_2} \Gamma^{N_1 N_2} \bigg) C* \epsilon
=0~,
\eea
\bea
\label{akse}
P_M \Gamma^M C* \epsilon +{1 \over 24} G_{N_1 N_2 N_3} \Gamma^{N_1 N_2 N_3} \epsilon =0~,
\eea
respectively, where $Q$ is a $U(1)$ connection of $\lambda$.

Evaluating the gravitino KSE on the near horizon fields (\ref{hormetr}), one finds that these can be integrated along the lightcone directions. In particular setting $\epsilon=\epsilon_++\epsilon_-$, $\Gamma_\pm \epsilon_\pm=0$, and solving the $-$ component of the gravitino KSE, one finds that
\bea
\label{st1}
\epsilon_+ = \phi_+~,~~~
\epsilon_-&=&\phi_- + r \Gamma_- \Theta_+ \phi_+~,
\eea
where $\phi_\pm$ do not depend on $r$, and  the solution of the  $+$ component of the gravitino KSE gives
\bea
\label{st2}
\phi_+ =\eta_+ + u \Gamma_+ \Theta_-\eta_-~,~~~\phi_- =\eta_-~,
\eea
where $\eta_\pm$ do not depend on both $u$ and $r$ coordinates, and where
\bea
\Theta_\pm= \bigg({1 \over 4} h_i \Gamma^i
\pm{i \over 12} Y_{n_1 n_2 n_3} \Gamma^{n_1 n_2 n_3} \bigg)
+\bigg( {1 \over 96} H_{\ell_1 \ell_2 \ell_3} \Gamma^{\ell_1 \ell_2 \ell_3} \pm {3 \over 16} \Phi_i \Gamma^i \bigg) C*~.
\eea

Furthermore, the integrability of the gravitino KSE along the lightcone directions requires the following algebraic conditions
\bea
\label{alg1}
\big({1 \over 2} \Delta -{1 \over 8} dh_{ij}\Gamma^{ij} \big)\phi_+ +{1 \over 16} L_{ij} \Gamma^{ij} C* \phi_+
+2 \Theta_- \tau_+
=0~,
\eea
\bea
\label{alg2}
(\Delta h_i - \partial_i \Delta) \Gamma^i \phi_+
+  \big(-{1 \over 2} dh_{ij} \Gamma^{ij} +{i \over 12} X_{\ell_1 \ell_2 \ell_3 \ell_4} \Gamma^{\ell_1
\ell_2 \ell_3 \ell_4} \big) \tau_+ +{1 \over 2}  L_{ij} \Gamma^{ij} C* \tau_+ =0~,
\nn
\eea
and
\bea
\label{extraalg}
\bigg(-{1 \over 2} \Delta -{1 \over 8} dh_{ij}\Gamma^{ij} +{i \over 48} X_{\ell_1 \ell_2 \ell_3 \ell_4}
\Gamma^{\ell_1 \ell_2 \ell_3 \ell_4} \bigg) \eta_- +{1 \over 8} L_{ij} \Gamma^{ij} C* \eta_-
-2 \Theta_+ \Theta_-   \eta_- =0~,
\eea
where we have set
\bea
\label{et}
\tau_+ = \Theta_+ \phi_+~.
\eea

Next, we substitute the Killing spinor (\ref{st1})  into the gravitino KSE and evaluate the resulting expression along the directions transverse to the light cone to find

\bea
\label{skse1}
&&\tn_i \phi_+ + \bigg(-{i \over 2} \Lambda_i -{1 \over 4} h_i
-{i \over 4} Y_{i \ell_1 \ell_2} \Gamma^{\ell_1 \ell_2}
+{i \over 12} \Gamma_i{}^{\ell_1 \ell_2 \ell_3} Y_{\ell_1 \ell_2 \ell_3} \bigg) \phi_+
\cr
&&+ \bigg({1 \over 16} \Gamma_i{}^j \Phi_j -{3 \over 16} \Phi_i
-{1 \over 96} \Gamma_i{}^{\ell_1 \ell_2 \ell_3} H_{\ell_1 \ell_2 \ell_3}
+{3 \over 32} H_{i \ell_1 \ell_2} \Gamma^{\ell_1 \ell_2} \bigg) C* \phi_+=0~,
\eea

\bea
\label{skse2}
\tn_i \tau_+ + \bigg(-{i \over 2} \Lambda_i -{3 \over 4} h_i
+{i \over 4} Y_{i \ell_1 \ell_2} \Gamma^{\ell_1 \ell_2}
-{i \over 12} \Gamma_i{}^{\ell_1 \ell_2 \ell_3} Y_{\ell_1 \ell_2 \ell_3} \bigg) \tau_+~~~~~~~~~~~~~~~~~~~~~~~~~&&
\cr
+ \bigg(-{1 \over 16} \Gamma_i{}^j \Phi_j +{3 \over 16} \Phi_i
-{1 \over 96} \Gamma_i{}^{\ell_1 \ell_2 \ell_3} H_{\ell_1 \ell_2 \ell_3}
+{3 \over 32} H_{i \ell_1 \ell_2} \Gamma^{\ell_1 \ell_2} \bigg) C* \tau_+ ~~~~~~~~~~~&&
\cr
+ \bigg(-{1 \over 4} dh_{ij} \Gamma^j -{i \over 12} X_{i \ell_1 \ell_2 \ell_3} \Gamma^{\ell_1 \ell_2 \ell_3} \bigg) \phi_+
+ \bigg( {1 \over 32} \Gamma_i{}^{\ell_1 \ell_2} L_{\ell_1 \ell_2}-{3 \over 16} L_{ij} \Gamma^j \bigg) C* \phi_+ =0~,&&
\eea
and
\bea
\label{skse1x}
&&\tn_i \eta_- + \bigg(-{i \over 2} \Lambda_i +{1 \over 4} h_i
+{i \over 4} Y_{i \ell_1 \ell_2} \Gamma^{\ell_1 \ell_2}
-{i \over 12} \Gamma_i{}^{\ell_1 \ell_2 \ell_3} Y_{\ell_1 \ell_2 \ell_3} \bigg) \eta_-
\cr
&&+ \bigg(-{1 \over 16} \Gamma_i{}^j \Phi_j +{3 \over 16} \Phi_i
-{1 \over 96} \Gamma_i{}^{\ell_1 \ell_2 \ell_3} H_{\ell_1 \ell_2 \ell_3}
+{3 \over 32} H_{i \ell_1 \ell_2} \Gamma^{\ell_1 \ell_2} \bigg) C* \eta_- =0~,
\eea
where $\Lambda$ is the restriction of $Q$ along ${\cal S}$. Note that $\Lambda$ is independent of $r,u$ because the near horizon scalars
do not depend on these coordinates.

It remains to evaluate the dilatino KSE ({\ref{akse}}) on the spinor (\ref{st1}). A direct substitution reveals that
\bea
\label{alg3}
\bigg(-{1 \over 4} \Phi_i \Gamma^i +{1 \over 24} H_{\ell_1 \ell_2 \ell_3} \Gamma^{\ell_1 \ell_2 \ell_3}
\bigg) \phi_+ + \xi_i \Gamma^i C* \phi_+ =0~,
\eea
\bea
\label{alg4}
\bigg( -{1 \over 4} \Phi_i \Gamma^i -{1 \over 24} H_{\ell_1 \ell_2 \ell_3} \Gamma^{\ell_1 \ell_2 \ell_3} \bigg) \tau_+
- \xi_i \Gamma^i C* \tau_+ +{1 \over 8} L_{ij} \Gamma^{ij} \phi_+ =0~,
\eea
and
\bea
\label{alg3x}
\bigg({1 \over 4} \Phi_i \Gamma^i +{1 \over 24} H_{\ell_1 \ell_2 \ell_3} \Gamma^{\ell_1 \ell_2 \ell_3}
\bigg) \eta_- + \xi_i \Gamma^i C* \eta_- =0~.
\eea

This concludes the evaluation of the KSEs on the IIB near horizon geometries and their integration along the lightcone directions.

\subsection{Independent horizon KSEs} \label{ihkse}

The conditions on the Killing spinors we have found in the previous section are not independent. It is customary in the investigation
of supersymmetric solutions that one first solves the KSEs.  Then one imposes those Bianchi identities and field equations that are
not implied as integrability conditions of the KSEs. Here, as in \cite{5index} and \cite{11index}, we shall adopt a different strategy.
We shall use the field equations and Bianchi identities to find which of the conditions implied by the KSEs presented in the previous section are independent.

To determine the independent KSEs on the horizon section ${\cal S}$ is more involved than similar results for M-horizons \cite{11index}.  Because
of this, we shall state here the result and the proof can be found in appendix B. In particular upon the use of the field equations and Bianchi identities
of IIB supergravity, the independent KSEs are
\bea
\nabla^{(\pm)}_i \eta_\pm\equiv \tilde \nabla_i\eta_\pm + \Psi^{(\pm)}_i \eta_\pm=0~,
\la{paral}
\eea
and
\bea
{\cal A}^{(\pm)}\eta_\pm=0~,
\la{alge}
\eea
where
\bea
\Psi^{(\pm)}_i &=&-{i \over 2} \Lambda_i \mp{1 \over 4} h_i
\mp {i \over 4} Y_{i \ell_1 \ell_2} \Gamma^{\ell_1 \ell_2}
\mp {i \over 12} \Gamma_i{}^{\ell_1 \ell_2 \ell_3} Y_{\ell_1 \ell_2 \ell_3}
\cr
&&+ \bigg(\pm{1 \over 16} \Gamma_i{}^j \Phi_j \mp {3 \over 16} \Phi_i
-{1 \over 96} \Gamma_i{}^{\ell_1 \ell_2 \ell_3} H_{\ell_1 \ell_2 \ell_3}
+{3 \over 32} H_{i \ell_1 \ell_2} \Gamma^{\ell_1 \ell_2} \bigg) C* ~,
\la{psipm}
\eea
and
\bea
{\cal A}^{(\pm)}=\mp {1 \over 4} \Phi_i \Gamma^i +{1 \over 24} H_{\ell_1 \ell_2 \ell_3} \Gamma^{\ell_1 \ell_2 \ell_3} + \xi_i \Gamma^i C* ~.
\eea
Furthermore if $\eta_-$ solves (\ref{paral}) and (\ref{alge}), then
\bea
\eta'_+=\Gamma_+ \Theta_-\eta_-~,
\la{lmap}
\eea
also solves (\ref{paral}) and (\ref{alge}).

Therefore the independent KSEs are those that one finds after a naive restriction of the KSEs of IIB supergravity
on ${\cal S}$ and after considering the lightcone projections of the Killing spinor. However, the additional property that $\eta_+'$ solves the KSEs does not arise in this way
and a more thorough analysis of the near horizon KSEs is required to establish this.

\newsection{Horizon Dirac equations}

We have seen that the gravitino KSE gives rise to two parallel transport equations on ${\cal S}$ associated with the covariant derivatives
$\nabla^{(\pm)}$  (\ref{paral}). If $S^\pm$ are the complex chiral spin bundles over ${\cal S}$, then $\nabla^{\pm}: \Gamma(S^\pm\otimes \lambda^{1\over2})\rightarrow \Gamma(S^\pm\otimes \lambda^{1\over2})$, where $\Gamma(S^\pm\otimes \lambda^{1\over2})$ are the smooth sections of $S^\pm\otimes \lambda^{1\over2}$. In turn,
one can define the associated Dirac operators
\bea
{\cal D}^{(\pm)}\equiv \Gamma^i \nabla^{(\pm)}_i=\Gamma^i \tilde\nabla_i+\Psi^\pm~,
\la{dirac}
\eea
where
\bea
\Psi^\pm\equiv \Gamma^i \Psi_i^{(\pm)}=-{i \over 2} \Lambda_i \Gamma^i \mp {1 \over 4} h_i \Gamma^i \pm {i \over 6} Y_{\ell_1 \ell_2 \ell_3} \Gamma^{\ell_1
\ell_2 \ell_3}+ \big( \pm {1 \over 4} \Phi_i \Gamma^i +{1 \over 24} H_{\ell_1 \ell_2 \ell_3} \Gamma^{\ell_1 \ell_2 \ell_3}\big) C*
\eea
Clearly the $\nabla^{\pm}$ parallel spinors   are  zero modes of ${\cal D}^{(\pm)}$. Here we shall prove the converse. In particular, we shall show that
all zero modes of the horizon Dirac equations ${\cal D}^{(\pm)}$ are Killing spinors, i.e.~they are parallel with respect to the $\nabla^{\pm}$ connections and ${\cal A}^{(\pm)}\eta_\pm=0$. Therefore, we
shall establish
\bea
\nabla^{(\pm)}\eta_\pm=0~,~~~{\cal A}^{(\pm)}\eta_\pm=0\Longleftrightarrow {\cal D}^{(\pm)}\eta_\pm=0~.
\la{lich}
\eea
In addition, we shall demonstrate
that
\bea
\parallel \eta_+\parallel=\mathrm{const.}
\la{const}
\eea
We shall prove the above statements separately for the ${\cal D}^{(\pm)}$  horizon operators.

\subsection{A maximum principle for $\parallel \eta_+\parallel^2$} \label{slich1}

As we have mentioned, if $\eta_+$ is Killing spinor then $\eta_+$ is a zero mode of ${\cal D}^{(+)}$ which demonstrates (\ref{lich}) in one direction. It remains to show the
 opposite direction of  (\ref{lich}), and (\ref{const}).  For this, we shall formulate a maximal principle for the scalar function $\parallel \eta_+\parallel^2$ of ${\cal S}$.
 In particular, assuming that ${\cal D}^{(+)}\eta_+=0$ and after some extensive Clifford algebra which is described in detail in appendix C, one establishes
 \bea
\tn^i \tn_i \parallel \eta_+\parallel^2
- h^i \tn_i \parallel \eta_+\parallel^2
= 2 \parallel \nabla^{(+)}\eta_+ \parallel^2+  \parallel{\cal{A}}^{(+)}\eta_+\parallel^2~.
\label{lichident}
\eea
 Since ${\cal S}$ is compact and the right hand side  of (\ref{lichident}) is positive, an application of the maximum principle implies that (\ref{const}) holds and that $\eta_+$ is a Killing spinor. This establishes both (\ref{const}) and the opposite direction in (\ref{lich}) for the $\eta_+$ spinors.

 We remark that $\parallel\eta_+\parallel = {\rm const}$ implies
 the following conditions
 \bea
 \label{extramatch2}
 - \Delta\, \parallel\eta_+\parallel^2 +4  \parallel\Theta_+ \eta_+\parallel^2 =0~,~~~\mathrm{Re}\,\langle \eta_+ , \Gamma_i \Theta_+ \eta_+ \rangle  =0 \ .
 \la{newcon}
\eea
Alternatively, these conditions can also be derived from the requirement that the vector bilinear in section 6 is Killing.

\subsection{A Lichnerowicz theorem for ${\cal D}^{(-)}$} \label{slich2}
Again if $\eta_-$ is a Killing spinor, then $\eta_-$ is a zero mode of the horizon Dirac operator ${\cal D}^{(-)}$ which establishes one direction in (\ref{lich}). To prove the converse, we use
\bea
\Gamma^{ij} {\tilde{\nabla}}_i {\tilde{\nabla}}_j \eta_- = -{1 \over 4} {\tilde{R}} \,\eta_- \ ,
\eea
to find after some extensive Clifford algebra, which is described in appendix C, the identity
\bea
\int_{\cal S}\, \parallel {\cal D}^{(-)}\eta_-\parallel^2=\int_{\cal S}\, \parallel \nabla^{(-)}\eta_-\parallel^2+ {1\over2} \int_{\cal S}\, \parallel {\cal A}^{(-)}\eta_-\parallel^2+
\int_{\cal S} \mathrm{Re} \langle {\cal B} \eta_-, {\cal D}^{(-)}\eta_-\rangle+ {\cdots}~,
\la{intid}
\eea
where ${\cal B}$ is a Clifford algebra element that depends on the fluxes, and the dots represent surface terms and terms that depend on the field equations and Bianchi identities, see appendix C. Clearly the surface  terms vanish
because ${\cal S}$ is compact without boundary.
It is evident from (\ref{intid}) that if $ {\cal D}^{(-)}\eta_-=0$ and the field equations and Bianchi identities of IIB supergravity are satisfied, then
$\eta_-$ is a Killing spinor. This establishes the opposite direction in (\ref{lich}) for the $\eta_-$ spinors.

\newsection{Index theorem and supersymmetry}

We shall now establish the relation (\ref{index}) between the number of supersymmetries $N$  preserved by the IIB horizons and the index of Dirac operator presented in the
introduction.  First observe that
\bea
N=N_++N_-~,
\la{nnn}
\eea
where
\bea
N_\pm=\mathrm{dim}\,\mathrm{Ker}(\nabla^{(\pm)},  {\cal A}^{(\pm)})~.
\la{nk}
\eea
On the other hand, the Lichnerowicz  type theorem established in (\ref{lich}) implies that
\bea
N_\pm=\mathrm{dim}\,\mathrm{Ker}({\cal D}^{(\pm)})~.
\eea

Next, to apply the index theorem, first observe that the spinor bundle over the spacetime twisted by $\lambda^{1\over2}$ when restricted on ${\cal S}$ decomposes
as $S\otimes \lambda^{1\over2}=S_+\otimes \lambda^{1\over2}\oplus S_-\otimes \lambda^{1\over2}$.  In addition
since the IIB spinors lie in the positive chirality Weyl representation of $Spin_c(9,1)$,  $S_+\otimes \lambda^{1\over2}$   can be identified with the positive chirality
 spinor bundle $\tilde S_+\otimes \lambda^{1\over2}$ of $Spin_c(8)$. On the other hand $S_-\otimes \lambda^{1\over2}$ can be identified with the negative chirality spinor bundle $\tilde S_-\otimes \lambda^{1\over2}$ of $Spin_c(8)$ because if $\a_-$ is a section\footnote{Note that the subscript in $\a_-$ denotes $Spin(8)$ chirality while the subscript in $\eta_-$
  denotes a $\Gamma_-$ projection.} of $\tilde S_-\otimes \lambda^{1\over2}$, then $\eta_-=\Gamma_-\a_-$ is a section of  $S_-\otimes \lambda^{1\over2}$ and vice versa.
Using these identifications of spinor bundles, one concludes that   ${\cal D}^{(+)}: \Gamma(\tilde S_+\otimes \lambda^{1\over2})\rightarrow
 \Gamma(\tilde S_-\otimes \lambda^{1\over2})$, where $ \Gamma(\tilde S_\pm\otimes \lambda^{1\over2})$ denotes the smooth section of these bundles. The horizon Dirac operator ${\cal D}^{(+)}$
 has the same principal symbol as a $\lambda^{1\over2}$ twisted Dirac operator ${\cal D}_\lambda$ and therefore the same index. As a result
 \bea
 \mathrm{Index} ({\cal D}^{(+)})=\mathrm{dim}\,\mathrm{Ker}({\cal D}^{(+)})-\mathrm{dim}\,\mathrm{Ker}(({\cal D}^{(+)})^\dagger)=2\, \mathrm{Index} ({\cal D}_\lambda)~.
 \la{infom}
\eea
The factor of 2 appears in the right hand side of (\ref{infom}) because we count the dimension of the index of ${\cal D}^{(+)}$ over the real numbers.

To proceed, we shall  show that
\bea
N_-=\mathrm{dim}\,\mathrm{Ker}(({\cal D}^{(+)})^\dagger)~.
\la{nminus}
\eea
For this observe that
\bea
({\cal D}^{(+)})^\dagger=-\Gamma^i \tilde\nabla_i+{i \over 2} \Lambda_i \Gamma^i - {1 \over 4} h_i \Gamma^i + {i \over 6} Y_{\ell_1 \ell_2 \ell_3} \Gamma^{\ell_1
\ell_2 \ell_3}+ \big( {1 \over 4} \Phi_i \Gamma^i -{1 \over 24} H_{\ell_1 \ell_2 \ell_3} \Gamma^{\ell_1 \ell_2 \ell_3}\big) C*~,
\eea
where the conjugate is taken with the real part of the hermitian inner product in $\tilde S_-\otimes \lambda^{1\over2}$,
and that
\bea
{\cal D}^{(-)}\Gamma_-=\Gamma_-({\cal D}^{(+)})^\dagger~.
\eea
Therefore the dimension of the  kernel of ${\cal D}^{(-)}$ is the same as that of $({\cal D}^{(+)})^\dagger$ which establishes (\ref{nminus}).
Next combining (\ref{nnn}), (\ref{infom}) and (\ref{nminus}), one can show (\ref{index}) in the introduction with $N_-$ given by (\ref{nk}).

\newsection{ $\eta_+$ from $\eta_-$ Killing spinors } \label{ch5}

In this section, we shall explore further the observation made in section \ref{ihkse} that if $\eta_-\not=0$ is a Killing spinor, then $\eta_+'=\Gamma_+\Theta_-\eta_-$ is also a Killing spinor.
In particular, we shall show  that
\bea
\mathrm{Ker}\, \Theta_-=\{0\}
\eea
for all horizons with fluxes, otherwise the  metric decomposes as a product $\bR^{1,1}\times X^8$, the holonomy of $X^8$ is a subgroup of $Spin(7)$ and all the fluxes vanish.
 Therefore for horizons with non-trivial fluxes, $\Gamma_+\Theta_-$ is an injection and so $\mathrm{Index} (D_\lambda)\geq 0$.

To show this  suppose that there is $\eta_-\not=0$ Killing spinor, i.e.~$\nabla^{(-)}\eta_-={\cal A}^{(-)}\eta_-=0$, such that
\bea
\Theta_-\eta_-=0~.
\label{van1}
\eea
To proceed, note that ({\ref{extraalg}}) together with ({\ref{van1}}) imply that
\bea
\bigg(-{1 \over 2} \Delta -{1 \over 8} dh_{ij} \Gamma^{ij} +{i \over 48} X_{\ell_1 \ell_2 \ell_3 \ell_4}
\Gamma^{\ell_1 \ell_2 \ell_3 \ell_4} \bigg) \eta_- +{1 \over 8} L_{ij} \Gamma^{ij} C* \eta_- =0 ~.
\eea
In turn this implies that
\bea
\Delta \langle \eta_- , \eta_- \rangle =0~,
\eea
where we have used the identity $\langle \eta_- , \Gamma_{ij} C* \eta_- \rangle =0$.
Since $\eta_-$ is nowhere vanishing,
\bea
\Delta =0 \ .
\eea
Next, using ({\ref{skse1x}}), we compute
\bea
\label{exdif1}
\tn_i \langle \eta_- , \eta_- \rangle &=& -{1 \over 2} h_i \langle \eta_- , \eta_- \rangle
+ \langle \eta_-, -{i \over 2} Y_{i \ell_1 \ell_2} \Gamma^{\ell_1 \ell_2} \eta_- \rangle
\nonumber \\
&+& 2 {\rm Re \ } \bigg( \langle \eta_-, \Gamma_i \big(-{3 \over 16} \Phi_j \Gamma^j
+{1 \over 96} H_{\ell_1 \ell_2 \ell_3} \Gamma^{\ell_1 \ell_2 \ell_3} \big) C* \eta_- \rangle~.
\eea
Then, on substituting ({\ref{van1}}) into ({\ref{exdif1}}) in order to eliminate the $\Phi$ and $H$-terms,
one obtains
\bea
\label{exdif2}
\tn_i \langle \eta_- , \eta_- \rangle &=& - h_i \langle \eta_- , \eta_- \rangle~.
\eea
 Then on using ({\ref{feq5}}), together with $\Delta=0$, ({\ref{exdif2}}) implies that
 \bea
 \tn^i \tn_i \langle \eta_-, \eta_- \rangle = \bigg({4 \over 3} Y_{\ell_1 \ell_2 \ell_3} Y^{\ell_1 \ell_2 \ell_3}
 +{3 \over 4} \Phi_i {\bar{\Phi}}^i +{1 \over 24} H_{\ell_1 \ell_2 \ell_3} {\bar{H}}^{\ell_1 \ell_2 \ell_3} \bigg)~.
 \langle \eta_- , \eta_- \rangle
 \eea
 On integrating both sides of this expression over ${\cal{S}}$, one finds that
 \bea
 Y=0, \quad H=0, \quad \Phi =0~.
 \eea
 Returning to ({\ref{feq5}}), we have
 \bea
 \tn^i h_i = h^2
 \eea
 and so integrating both sides of this expression over ${\cal{S}}$ one finds
 \bea
 h=0
 \eea
 as well.
With these conditions, it is straightforward to see that the Bianchi identities also imply that
\bea
X=0, \quad L=0 \ .
\eea
Hence, the 5-form $F$ and complex 3-form $G$ vanish.

 It remains to consider the complex 1-form $\xi$, and the connection $\Lambda$.
The spacetime is $\bR^{1,1} \times {\cal{S}}$, where ${\cal{S}}$ admits a spinor $\eta_-$ satisfying
 \bea
 \label{kah}
 \tn_i \eta_- = {i \over 2} \Lambda_i \eta_-~,
 \eea
 whose Ricci tensor is given as
 \bea
 {\tilde{R}}_{ij} = 2 \xi_{(i} {\bar{\xi}}_{j)}~,
 \eea
 and the algebraic KSE ({\ref{alg3x}}) reduces to
 \bea
 \label{simpleralg}
 {\bar{\xi}}_i \Gamma^i \eta_- =0~.
 \eea
 We also have
 \bea
 \tn^i \xi_i -2i \Lambda^i \xi_i =0~,
 \eea
 and
 \bea
 d \Lambda = -i \xi \wedge {\bar{\xi}} \ .
 \eea
Note that ({\ref{kah}}) implies that ${\cal{S}}$ is K\"ahler{\footnote{This can be shown by observing that 2-form spinor bilinear constructed
from $\eta_-$ is not degenerate and it is covariantly constant with respect to the Levi-Civita connection.}}.
 There are two cases to consider, corresponding as to whether $\eta_-$ is a pure spinor, or not.

First, observe that ({\ref{simpleralg}}) implies that either $\eta_-$ is a pure spinor or $\xi=0$. If $\xi=0$, ${\cal S}$ is Ricci flat and $d\Lambda=0$. Thus ${\cal S}$
 up to a finite cover has holonomy contained in $Spin(7)$.

Next, suppose that $\eta_-$ is pure.
Then ({\ref{simpleralg}}) implies that
\bea
\xi^i \xi_i =0~.
\eea
Using this, the bosonic field equations and Bianchi identities, together with $\xi^i \xi_i=0$, one can show  that
\bea
\tn_i \tn^i \big( \xi^j {\bar{\xi}}_j \big)= 2 \bigg( \tn_{(i} \xi_{j)}-2i \Lambda_{(i} \xi_{j)} \bigg) \bigg(\tn^{(i} {\bar{\xi}}^{j)}
+2i \Lambda^{(i} {\bar{\xi}}^{j)} \bigg) + 6 \xi^i {\bar{\xi}}_i \xi^j {\bar{\xi}}_j~.
\eea
On applying the maximum principle, one finds\footnote{The application of the maximum principle requires that $\xi$ is at least $C^2$  differentiable. So  our results imply that the
cosmic string solutions of \cite{yau} and   the
 D7 branes of \cite{gibbonsd7} with compact transverse space, including those with 24 strings and 24 D7-branes, respectively, cannot be more than $C^1$ differentiable.}
\bea
\xi =0~.
\eea
Therefore all the fluxes vanish and establishes our result.

\section{The dynamical $\mathfrak{sl}(2,\mathbb{R})$ symmetry of IIB horizons}

\subsection{Killing vectors}

 IIB horizons with $N_-=0$ coincide with those for which the bi-linear matching condition has been imposed and their geometry has already been investigated in \cite{iibhor}. Here we shall
 explore some aspects of the geometry of horizons with $N_-\not=0$. In particular,
 we shall demonstrate that if $N_-\not=0$, then the IIB horizons with non-trivial fluxes admit an $\mathfrak{sl}(2,\bR)$ symmetry subalgebra.
First note that the IIB horizons (\ref{hormetr})  are invariant under the symmetries generated by the vector fields  $\partial_u$ and $u\partial_u-r\partial_r$.  Here we shall show
that they admit an additional symmetry which enhances the symmetry algebra to  $\mathfrak{sl}(2,\mathbb{R})$. Such an additional symmetry is a consequence of the supersymmetry
  enhancement that we have already demonstrated for IIB horizons with non-trivial fluxes and $N_-\not=0$. Both the additional supersymmetry and  $\mathfrak{sl}(2,\mathbb{R})$ symmetry are dynamical
as they arise as a consequence of the IIB field equations.

Using (\ref{st1}) and (\ref{st2}), one finds that the most general Killing spinor is
\begin{eqnarray}
\epsilon=\eta_++ u \Gamma_+\Theta_-\eta_-+\eta_-+r \Gamma_-\Theta_+\eta_++ r u \,\Gamma_-\Theta_+\Gamma_+\Theta_-\eta_-~.
\la{ksee}
\end{eqnarray}
Since we have assumed that $N_-\not=0$, there is an $\eta_-\not=0$ which solves the KSEs $\nabla^{(-)}\eta_-={\cal A}^{(-)}\eta_-=0$ on the horizon section ${\cal S}$.  Furthermore,
 since we have also assumed that the horizon does not have trivial fluxes $\Theta_-$ is an injection and so there is a spinor $\eta_+=\Gamma_+\Theta_-\eta_-$ which also solves
 the KSEs on the horizon section ${\cal S}$, i.e.~$\nabla^{(+)}\eta_+={\cal A}^{(+)}\eta_+=0$.  Since $\eta_-$ and $\eta_+$ are linearly independent, they give
 rise to two Killing spinors (\ref{ksee})  which can be constructed from the pairs $(\eta_-,0)$ and $(\eta_-, \eta_+)$. After a rearrangement, the two  Killing spinors (\ref{ksee})
 can be written as
\begin{eqnarray}
\epsilon_1=\eta_-+u \eta_++ru\Gamma_-\Theta_+\eta_+ ~,~~~\epsilon_2=\eta_++ r \Gamma_- \Theta_+\eta_+~,~~~\eta_+=\Gamma_+\Theta_-\eta_-~.
\label{tks}
\end{eqnarray}

To continue, we shall use the property of the KSEs of IIB supergravity that if  $\zeta_1$ and $\zeta_2$ are Killing spinors, then the 1-form bilinear\footnote{The 1-form bilinear which gives rise to a Killing vector is that associated with the real part of the Dirac inner product of  $Spin(9,1)$, see \cite{iibspingeom}. }
\begin{eqnarray}
K=\mathrm{Re}\, \langle(\Gamma_+-\Gamma_-)\zeta_1, \Gamma_A \zeta_2\rangle \, e^A~,
\label{bil}
\end{eqnarray}
is associated with a Killing vector which
also preserves the 1-, 3- and 5-form fluxes \cite{iibspingeom, figueroa}, see appendix D for a short proof of this statement. In particular, from the two Killing spinors
(\ref{tks}), one can construct three 1-form bi-linears.  A   substitution of (\ref{tks}) into (\ref{bil}) reveals
\begin{eqnarray}
 K_1=\mathrm{Re}\,\langle(\Gamma_+-\Gamma_-)\epsilon_1, \Gamma_A \epsilon_2\rangle \, e^A&=& (2r\, \mathrm{Re}\,\langle\Gamma_+\eta_-, \Theta_+\eta_+\rangle+   u r^2 \Delta \parallel\eta_+\parallel^2) \,{\bf{e}}^+
 \nonumber \\
 &-&2u \parallel\eta_+\parallel^2\, {\bf{e}}^-  + V_i {\bf{e}}^i~,
 \nonumber \\
 K_2=\mathrm{Re}\,\langle(\Gamma_+-\Gamma_-)\epsilon_2, \Gamma_A \epsilon_2\rangle \, e^A &=& r^2 \Delta \parallel\eta_+\parallel^2 {\bf{e}}^+-2 \parallel\eta_+\parallel^2 {\bf{e}}^-
\nonumber \\
K_3=\mathrm{Re}\,\langle(\Gamma_+-\Gamma_-)\epsilon_1, \Gamma_A \epsilon_1\rangle \, e^A&=&(2\parallel\eta_-\parallel^2+4r u \mathrm{Re}\,\langle\Gamma_+\eta_-, \Theta_+\eta_+\rangle
 +  r^2 u^2 \Delta \parallel\eta_+\parallel^2) {\bf{e}}^+
\nonumber \\
 &-&2u^2 \parallel\eta_+\parallel^2 {\bf{e}}^- +2u V_i {\bf{e}}^i~,
 \nonumber \\
 \label{b1forms}
 \end{eqnarray}
where we have set
\begin{eqnarray}
\label{extraiso}
V_i = \mathrm{Re}\, \langle \Gamma_+ \eta_- , \Gamma_i \eta_+ \rangle~ \ ,
\end{eqnarray}
and we have used ({\ref{extramatch2}}) to simplify the components of the $K_a$, $a=1,2,3$.
By construction, all three 1-forms must give rise to symmetries of the IIB background.  In particular, one has that
\bea
{\cal L}_{K_a} g=0~,~~~{\cal L}_{K_a} F=0~,~~~ {\cal L}_{K_a} P=2i\,  Q_a P~,~~~{\cal L}_{K_a} G=i\, Q_a G~,~~~{\cal L}_{K_a} dQ=0~,
\la{incon}
\eea
where $Q_a=i_{K_a} Q$. Observe that the $P$ and $G$ fluxes are invariant up to a $U(1)$ transformation, and we shall assume that  $dQ$
is equivariant under the group action.
To find the conditions that this imposes on the geometry
of ${\cal S}$, we shall investigate two different cases depending on whether $V=0$ or not.

\subsection{$V\not=0$}

To find the restrictions on the geometry of ${\cal S}$ and the fluxes imposed by  $K_1, K_2$ and $K_3$, we decompose the conditions (\ref{incon})
 along the lightcone and transverse directions and after some straightforward computation, we find that
\begin{eqnarray}
\tilde\nabla_{(i} V_{j)}=0~,~~~\tilde {\cal L}_Vh=0 ~,~~~\tilde {\cal L}_V\Delta=0~,~~~\tilde {\cal L}_V Y=0~,
\cr
\tilde {\cal L}_V L=i \Lambda_V L~,~~\tilde {\cal L}_V \Phi=i   \Lambda_V \Phi~,~~\tilde {\cal L}_V H=i  \Lambda_V H~,~~\tilde {\cal L}_V \xi=2i  \Lambda_V\xi~,~~{\cal L}_V d\Lambda=0~,
\end{eqnarray}
where $\Lambda_V=i_V \Lambda$.
Therefore, ${\cal S}$ admits an isometry generated by $V$ which leaves $h, \Delta, Y$, and transforms $L, \Phi, H$ and $\xi$ up to a $U(1)$ transformation.
In addition, one finds the conditions
\begin{eqnarray}
&&-2 \parallel\eta_+\parallel^2-h_i V^i+2 \mathrm{Re}\,\langle\Gamma_+\eta_-, \Theta_+\eta_+\rangle=0~,~~~i_V (dh)+2 d \mathrm{Re}\,\langle\Gamma_+\eta_-, \Theta_+\eta_+\rangle=0~,
\cr
&& 2 \mathrm{Re}\, \langle\Gamma_+\eta_-, \Theta_+\eta_+\rangle-\Delta \parallel\eta_-\parallel^2=0~,~~~
V+ \parallel\eta_-\parallel^2 h+d \parallel\eta_-\parallel^2=0~.
\label{concon}
\end{eqnarray}
These follow from the KSEs and the field equations, and we shall use them later to further simplify the vector fields associated with $K_1$, $K_2$ and $K_3$.
Notice that the last equality in (\ref{concon}) expresses $V$ in terms of $h$.  A similar relation has been derived
 for heterotic and M horizons \cite{11index}, \cite{hhor}. Furthermore, one has that
\begin{eqnarray}
{\cal L}_V\parallel\eta_-\parallel^2=0~.
\label{inphim}
\end{eqnarray}

It is likely that in addition to the Killing vectors associated with $K_1$, $K_2$ and $K_3$, there are additional conditions on the geometry of ${\cal S}$. We shall not
elaborate on these here. The results will be reported elsewhere.

\subsubsection{$V=0$}

A special case arises whenever $V=0$. In this case, the group action generated by $K_1, K_2$ and $K_3$ has only 2-dimensional orbits. A direct substitution of this condition in (\ref{concon}) reveals that
\begin{eqnarray}
\Delta \parallel\eta_-\parallel^2=2 \parallel\eta_+\parallel^2~,~~~h=\Delta^{-1} d\Delta~.
\la{vzdelta}
\end{eqnarray}
Since $dh=0$, and $h$ is exact, such horizons are static. Using this, the field equation (\ref{feq8}) implies that
 \bea
 X=L=0~.
 \eea
 After a coordinate transformation $r\rightarrow \Delta r$, the near horizon geometry becomes a warped product of $AdS_2$ with ${\cal S}$, $AdS_2\times_w {\cal S}$. Therefore,
 in this way one can identify the most general $AdS_2$ backgrounds of IIB supergravity. As a consequence of our results, IIB $AdS_2$ backgrounds  preserve at least 2 supersymmetries.

\subsection{$\mathfrak{sl}(2,\mathbb{R})$ symmetry of  IIB horizons}

It remains to show that all IIB horizons with non-trivial fluxes and $N_-\not=0$ admit an $\mathfrak{sl}(2,\mathbb{R})$ symmetry. For this,  we use the various identities derived in the
previous section (\ref{concon}) to write the vector fields associated to the 1-forms $K_1, K_2$ and $K_3$ (\ref{b1forms}) as
\begin{eqnarray}
K_1&=&-2u \parallel\eta_+\parallel^2 \partial_u+ 2r \parallel\eta_+\parallel^2 \partial_r+ V^i \tilde \partial_i~,
\cr
K_2&=&-2 \parallel\eta_+\parallel^2 \partial_u~,
\cr
K_3&=&-2u^2 \parallel\eta_+\parallel^2 \partial_u +(2 \parallel\eta_-\parallel^2+ 4ru \parallel\eta_+\parallel^2)\partial_r+ 2u V^i \tilde \partial_i~,
\end{eqnarray}
where we have used the same symbol for the 1-forms and the associated vector fields. A direct computation then reveals using (\ref{inphim}) that
\begin{eqnarray}
[K_1,K_2]=2 \parallel\eta_+\parallel^2 K_2~,~~~[K_2, K_3]=-4 \parallel\eta_+\parallel^2 K_1  ~,~~~[K_3,K_1]=2 \parallel\eta_+\parallel^2 K_3~. \ \ \ \
\end{eqnarray}
Therefore all such  IIB horizons with non-trivial fluxes admit an $\mathfrak{sl}(2,\mathbb{R})$ symmetry subalgebra. Note also that the orbits of the vector fields
are either 2- or 3-dimensional depending on whether $V=0$ or not.  As we have seen in the $V=0$ case, the orbits are $AdS_2$.
Furthermore,  if the fluxes are trivial, the spacetime is isometric to $\mathbb{R}^2\times {\cal S}$ and ${\cal S}$ admits at least one isometry. The symmetry group in this
case has an $\mathfrak{so}(1,1)\oplus \mathfrak{u}(1)$ subalgebra.

\newsection{Common sector horizons, and horizons with 5-form fluxes}

\subsection{5-form flux horizons}

IIB horizons with only 5-form fluxes have been investigated before in \cite{iibhorf5} using the bi-linear matching condition which requires that $\eta_-=0$.  Therefore  the horizons
that have been investigated so far are those with $N_-=0$, and it has been found that the near horizon sections include manifolds with a 2-SCYT structure. Since $N_-=0$,
the formula (\ref{index}) implies  $N=2\,\mathrm{Index}(D)$ and so the number of supersymmetries is a topological invariant. Note that for all these horizons
the line bundle $\lambda$ associated with the IIB scalars is trivial.

Next we shall assume that $N_-\not=0$. As we have already shown all such horizons with non-trivial fluxes admit an $\mathfrak{sl}(2,\bR)$ symmetry. Here we shall show that under an additional
assumption on $V$ all such horizons have vanishing Dirac index and admit
\bea
N=4k~,
\eea
supersymmetries.

To see this, we shall consider two separate cases. First take $V\not=0$. Assuming that the isometries
 generated by $V$ can be integrated to a circle action on ${\cal S}$, one can show \cite{atiyah2} that $\hat A [{\cal S}]=0$ and so the index of the Dirac operator vanishes.
As a result we have that $N_+=N_-$.  Since the KSEs of IIB supergravity  for backgrounds with only 5-form fluxes are linear over the complex numbers, $N_-$ is even and
so such horizons preserve 4k supersymmetries, $N_-=2k$.

Next suppose that $V=0$. In this case, we have shown that the horizons with non-trivial fluxes are warped products $AdS_2\times_w {\cal S}$. Observe that (\ref{vzdelta})  implies that $\Delta$ is no-where vanishing, and using $h=\Delta^{-1} d\Delta$,
the Einstein equation of ${\cal S}$ implies that the metric
\bea
g'=\Delta^{-{1\over7}} g
\eea
 has positive Ricci scalar, $\tilde R(g')>0$.  As a result, the index of the Dirac operator vanishes
and so again such horizons preserve $4k$ supersymmetries. Clearly in this case, there is a topological obstruction for ${\cal S}$ to be a solution of the KSEs which is the vanishing
of $\hat A$-genus.

The field equations impose additional conditions on $AdS_2\times_w {\cal S}$ horizons.  For example, they imply that $Y$ is a harmonic 3-form on ${\cal S}$. The solution
of the field equations and the geometry of ${\cal S}$ will be explored elsewhere.

\subsection{Common sector}

Another consistent truncation of IIB supergravity is to the common sector. The investigation of common sector horizons and their relation to the heterotic
ones \cite{hhor} have already been explored in \cite{iibhor} using the bilinear matching condition  which sets $\eta_-=0$.
Here we shall demonstrate that the bilinear matching condition does not impose a restriction  on common sector horizons and the investigation of their geometry reduces to that of the heterotic horizons described in \cite{hhor}.  To see this, it is convenient to work in the string frame\footnote{The relation of the IIB horizon fields to the string frame
  common sector horizon fields can be found in \cite{iibhor}. From here on all the fields are those of the common sector.}.   It is well known that in this frame, the
KSEs of IIB supergravity factorize to two pairs\footnote{We have used the signs $\pm$ in \cite{iibhor} to denote these two pairs of KSEs. We have changed the notation here
 to avoid confusion with the labeling of lightcone projections.} of KSEs depending on the string frame metric, the 3-form field strength ${\cal H}$, $d{\cal H}=0$, and the dilaton $\phi$, as
\bea
\hat\nabla \hat\e=0~,~~~~\hat{\cal  A}\hat\e=0~,
\cr
\check\nabla \check\e=0~,~~~~\check{\cal  A} \check\e=0~,
\la{tpairs}
 \eea
where  $\hat\nabla=\nabla+{1\over2} {\cal H}$ and  $\hat {\cal A}=\Gamma^M \partial_M\phi-{1\over12} {\cal H}_{MNR} \Gamma^{MNR}$, and
 $\check\nabla$ and $\check {\cal A}$ can be derived from these after setting ${\cal H}$ to $-{\cal H}$.

To continue, we can apply our results to one of two  pairs of KSEs above. In fact it is instructive to repeat the calculation we have done for IIB for the common sector in the string frame.
The common sector horizon metric is as in (\ref{hormetr}) and using $d{\cal H}=0$, one can write
\bea
{\cal H}=\bbe^+ \wedge \bbe^-\wedge S+ r \bbe^+ \wedge d_hS+\tilde {\cal H}~,~~~d\tilde{\cal H}=0,
\eea
where $S$ and $\tilde {\cal H}$ are a 1-form and a 3-form on ${\cal S}$, respectively.
Without loss of generality consider the first pair of KSEs in (\ref{tpairs}). Solving along the light cone directions, one finds
\bea
\label{cst1}
\hat\epsilon_+ = \hat\phi_+~,~~~
\hat\epsilon_-&=&\hat\phi_- + r \Gamma_- \Theta_+ \hat\phi_+~,
\eea
where $\hat\phi_\pm$ do not depend on $r$, and
\bea
\label{cst2}
\hat\phi_+ =\hat\eta_+ + u \Gamma_+ \Theta_-\hat\eta_-~,~~~\hat\phi_- =\hat\eta_-~,
\eea
where $\hat\eta_\pm$ do not depend on both $r$ and $u$, and
\bea
\Theta_\pm={1\over4} h_i \Gamma^i\mp {1\over 4} S_i \Gamma^i~.
\eea
An application of our IIB results implies that the remaining independent KSEs are
\bea
\hat{\tilde \nabla}_i \hat \eta_\pm\mp{1\over 4} h_i \hat\eta_\pm\pm{1\over4} S_i\hat\eta_\pm=0~,
\cr
\tilde\partial_i\phi \Gamma^i\hat\eta_\pm \pm{1\over2} S_i \Gamma^i\hat\eta_\pm-{1\over12} \tilde{\cal H}_{ijk} \Gamma^{ijk} \hat\eta_\pm=0~.
\la{ckses}
\eea

The analysis can proceed as in the general IIB case.
In particular, for horizons with non-trivial fluxes, one has $\mathrm{Ker}\, \Theta_-=\{0\}$.
To establish this, we make use of the following
common sector field equations:
\bea
\label{com1}
{\tilde{\nabla}}^i h_i = 2 \Delta +h^2-S^2
+2 h^i {\tilde{\nabla}}_i \phi \ ,
\eea
\bea
\label{com2}
{\tilde{\nabla}}^i S_i -2 S^i {\tilde{\nabla}}_i \phi =0 \ ,
\eea
and
\bea
\label{com3}
{\tilde{\nabla}}^i {\tilde{\nabla}}_i \phi
= h^i {\tilde{\nabla}}_i \phi +2 {\tilde{\nabla}}^i \phi
{\tilde{\nabla}}_i \phi +{1 \over 2}S^2 -{1 \over 12}
{\tilde{\cal{H}}}_{ijk} {\tilde{\cal{H}}}^{ijk} \ .
\eea
These conditions are obtained from the $+-$ components of
the Einstein, 3-form gauge, and the dilaton field equations.

Then, if $\mathrm{Ker}\, \Theta_- \neq \{0\}$, one obtains
\bea
S = -h
\eea
and furthermore,
\bea
\Delta=0
\eea
as a consequence of one of the algebraic conditions
obtained from the KSE. In addition, the common sector field equations ({\ref{com1}}), ({\ref{com2}}) and
({\ref{com3}}), together with
({\ref{ckses}}), imply the identity
\bea
{\tilde{\nabla}}^i {\tilde{\nabla}}_i
\bigg( e^{-2 \phi} \parallel {\hat{\eta}}_- \parallel^2 \bigg)
= {1 \over 6}  e^{-2 \phi} \parallel {\hat{\eta}}_- \parallel^2
{\tilde{\cal{H}}}_{ijk} {\tilde{\cal{H}}}^{ijk} \ .
\eea
Applying the maximum principle, one finds
that ${\cal{H}}=0$, and
\bea
{\tilde{\nabla}}_i \bigg( e^{-2 \phi} \parallel {\hat{\eta}}_- \parallel^2 \bigg)=0~,
\eea
which, together with ({\ref{ckses}}) also implies
\bea
h = -2 d \phi \ .
\eea
On substituting this condition into ({\ref{com1}}), an
additional
application of the maximum principle implies that
$\phi$ is constant, and $h=0$. So, if
$\mathrm{Ker}\, \Theta_- \neq \{0\}$ then all the fluxes vanish.

Next, one can show that
if $\hat\eta_-$ solves the KSEs (\ref{ckses}), then
\bea
\hat\eta_+=\Gamma_+ \Theta_- \hat\eta_-~,
\label{haa}
\eea
also solves (\ref{ckses}), using the same type of reasoning as in the IIB analysis.
The number of Killing spinors of common sector horizons is $N= \hat N+\check N$ with $\hat N=\hat N_++\hat N_-$, where $\hat N_\pm$ is the number of $\hat\eta_\pm$ Killing spinors, and
similarly for $\check N$. We have therefore shown that, if
${\hat{N}} \neq 0$, then ${\hat{N}}_+ \neq 0$.

To proceed further, the field equations ({\ref{com1}}),
({\ref{com2}}) and ({\ref{com3}}), together with
({\ref{ckses}}) imply that
\bea
{\tilde{\nabla}}^i {\tilde{\nabla}}_i \bigg( \parallel {\hat{\eta}}_+ \parallel^2 \bigg)
-(2 {\tilde{\nabla}}^i \phi + h^i)
{\tilde{\nabla}}_i \bigg( \parallel {\hat{\eta}}_+ \parallel^2 \bigg) =0~,
\eea
where we have also used the condition
\bea
\label{comalg2}
\parallel {\hat{\eta}}_+ \parallel^2
\bigg(\Delta - {1 \over 4}S^2+{1 \over 4}h^2 \bigg)=0~,
\eea
which follows from the KSE. An application of the maximum principle gives that
\bea
{\tilde{\nabla}}_i \bigg( \parallel {\hat{\eta}}_+ \parallel^2 \bigg) =0 \ .
\eea
Then ({\ref{ckses}}) and ({\ref{comalg2}}) imply that
\bea
S=h, \qquad \Delta =0 \ .
\eea

The conditions $S=h$ and $\Delta=0$ are the same as those  explored in \cite{hhor} for heterotic horizons.
So the common sector horizons can be investigated as special cases of the heterotic ones in \cite{hhor}.
In particular, one finds that
\bea
{\hat{\tilde{\nabla}}} h=0~,
\eea
so $h$ generates an isometry on ${\cal{S}}$,
and $h^2$ is constant. If $h=0$ then all the fluxes vanish.

We have proven that ${\hat{N}}_+ \neq 0$ if ${\hat{N}} \neq 0$. It is straightforward to show that ${\hat{N}}_- \neq 0$
, if ${\hat{N}} \neq 0$, as well. This is because, as was noted in \cite{hhor},
if $\hat\eta_+$ solves the KSEs, then
\bea
\hat \eta_-=\Gamma_- h_i \Gamma^i\hat \eta_+~,
\eea
also solves the KSEs, so $\hat N_-\not=0$  and the horizons preserve at least two supersymmetries.

In addition, a consequence of our analysis above is that the index of the Dirac  operator on
all common sector horizon sections vanishes. This is because $\Theta_-={1\over2}h_i\Gamma^i$ has an inverse\footnote{We have not been able to establish
 a similar property for all IIB horizons.} and pairs the zero modes of the $\Gamma^i\hat{\tilde \nabla}_i$ operator
and its adjoint. On all common sector sections, the $\hat A$-genus vanishes. This can also be seen from the property of horizon sections to admit an isometry generated by $h$.
An exhaustive analysis of the geometry of common sector horizons with extended supersymmetry can be done
using the method of \cite{hhor} applied for heterotic horizons.  As a result all common sector horizons preserve an even number of supersymmetries, as $\hat N=\hat N_++\hat N_-=2\hat N_-$ and similarly for $\check N$, and from the
classification results of \cite{hetclass}, if
$\hat N_->8$, then they are maximally supersymmetric with horizon sections  isometric to $T^8$.

\newsection{Conclusions}

We have demonstrated that the number of supersymmetries $N$ preserved by IIB black hole horizons can be expressed in terms of the index of the Dirac
operator on the horizon sections as in (\ref{index}). As a consequence of this formula, IIB horizons preserve an even number of supersymmetries. Moreover
 if  $N_-\not=0$, the horizons with non-trivial fluxes  exhibit an $\mathfrak{sl}(2,\bR)$ symmetry subalgebra. Furthermore, if the orbit of $\mathfrak{sl}(2, \bR)$ is 2-dimensional, then
 all such IIB horizons with non-trvial fluxes are warped products $AdS_2\times_w {\cal S}$. The proof of these results is based on the smoothness
of the horizons, the compactness of the horizon sections, as well as the demonstration of Lichnerowicz type  theorems which relate the Killing spinors
to the zero modes of horizon Dirac operators. Instrumental in the proof are the field equations and Bianchi identities of the theory, and so the symmetry enhancement exhibited
is dynamical. In addition, if $N_-=0$, the number of supersymmetries preserved by the horizons is given by the index of a Dirac operator, and so it is a topological invariant of
the horizon sections.
As a result, one can a priori test whether a manifold ${\cal S}$ with given Pontryagin numbers can admit a metric and fluxes such that it can identified
as a IIB horizon section preserving a given number $N$  of  supersymmetries.

The expression (\ref{index}) for the number of supersymmetries $N$ of IIB black hole horizons also applies to M-horizons and horizons of the 5-dimensional minimal gauged supergravity as
in the last two cases the index of the associated analogous horizon Dirac operator vanishes.   The similarities in the proof of (\ref{index}) for
all horizons so far, including IIB and M-horizons, suggests that this formula is universal and applies
to all supergravity theories. This is further supported by the observation that the IIB
 KSEs have a structure that encompasses that of the KSEs of all other supergravity theories, i.e.~it has both  a parallel transport equation associated
with gravitino supersymmetry transformation and an  algebraic KSE associated with
the supersymmetry transformation of the dilatino. Therefore for $D\geq 4$  supergravity theories with standard matter couplings, we shall propose the following.

\begin{itemize}

\item The number of supersymmetries preserved by supergravity horizons is given by
\bea
N=2N_-+  \mathrm{index}(D_E)
\eea
where $N_-$ is the dimension of the Kernel of a horizon Dirac operator which depends on the fluxes, and $D_E$ is the Dirac operator defined on an appropriate spinor bundle on the horizon sections ${\cal S}$ and twisted with $E$, and where $E$ is an appropriate vector bundle associated with the internal symmetries of the supergravity theory.

\item All supergravity horizons with $N_-\not=0$ and  non-trivial fluxes admit an $\mathfrak{sl}(2, \bR)$ symmetry subalgebra. Furthermore, if
the orbit of $\mathfrak{sl}(2, \bR)$ is 2-dimensional, then they are warped $AdS_2$ products.

\end{itemize}

It is not a priori apparent that $\mathrm{index}(D_E)$ will be an even number but in all examples investigated so far the index is either an even number or it vanishes.
Since the index vanishes on odd-dimensional manifolds, the proposed formula implies that all odd-dimensional supergravity horizons preserve an even number of supersymmetries, and if they have non-trivial fluxes it is likely that they will admit an $\mathfrak{sl}(2,\bR)$ symmetry
subalgebra. It is also expected that the index vanishes for non-chiral even-dimensional supergravities and so again the associated horizons preserve even number of supersymmetries
and admit an $\mathfrak{sl}(2,\bR)$ symmetry
subalgebra.

Furthermore, the existence of an $\mathfrak{sl}(2, \bR)$ symmetry subalgebra of supergravity horizons is closely related to the presence of non-trivial  fluxes.
Therefore the existence of the $\mathfrak{sl}(2, \bR)$ symmetry  is a property of the supergravity, and consequently a property of the low
energy approximation of string theory and M-theory. It is also an indication that in the context of the AdS/CFT correspondence all
such horizons have a conformal field theory dual.

\vskip 0.5cm
\noindent{\bf Acknowledgements} \vskip 0.1cm
\noindent  UG is supported by the Knut and Alice Wallenberg Foundation.
JG is supported by the STFC grant, ST/1004874/1.
GP is partially supported by the  STFC rolling grant ST/J002798/1.
\vskip 0.5cm

\setcounter{section}{0}\setcounter{equation}{0}

\appendix{Horizon Bianchi identities and field equations}

The Bianchi identities of IIB supergravity imply that
\bea
X = d_hY -{i \over 8} (\Phi \wedge {\bar{H}} - {\bar{\Phi}} \wedge H)~,~~~L = d_h \Phi  -i \Lambda \wedge \Phi + \xi \wedge {\bar{\Phi}}~.
\eea
The self-duality of $F$ requires that
\bea
X=-*_8 X~.
\la{sd1}
\eea
The remaining Bianchi identities are
\bea
d \star_8 Y &=& {i \over 8} H \wedge {\bar{H}}~,~~~dH = i \Lambda \wedge H - \xi \wedge {\bar{H}}~, ~~~
\cr
d \xi &=& 2i \Lambda \wedge \xi~,~~~d \Lambda = -i \xi \wedge {\bar{\xi}}~,
\la{ba}
\eea
where $\Lambda$ is a $U(1)$ connection of $\lambda$, see \cite{iibhor} for more details.

The independent field equations of IIB horizons are
\bea
\label{feq1}
\tn^i \Phi_i -i \Lambda^i \Phi_i - \xi^i {\bar{\Phi}}_i +{2 i \over 3} Y_{\ell_1 \ell_2 \ell_3} H^{\ell_1 \ell_2 \ell_3}=0~,
\eea

\bea
\label{feq3}
\tn^\ell H_{\ell ij} -i \Lambda^\ell H_{\ell ij}- h^\ell H_{\ell ij} + L_{ij} - \xi^\ell {\bar{H}}_{\ell ij}
+{2i \over 3}( \star_8 Y_{ij \ell_1 \ell_2 \ell_3} H^{\ell_1 \ell_2 \ell_3} - 6 Y_{ij \ell} \Phi^\ell) =0~,
\eea

\bea
\label{feq4}
\tn^i \xi_i -2i \Lambda^i \xi_i - h^i \xi_i +{1 \over 24}(-6 \Phi^i \Phi_i + H_{\ell_1 \ell_2 \ell_3} H^{\ell_1 \ell_2 \ell_3}) =0~,
\ee

\bea
\label{feq5}
{1 \over 2} \tn^i h_i - \Delta - {1 \over 2} h^2 + {2 \over 3} Y_{\ell_1 \ell_2 \ell_3} Y^{\ell_1 \ell_2 \ell_3}
+{3 \over 8} \Phi^i {\bar{\Phi}}_i +{1 \over 48} H_{\ell_1 \ell_2 \ell_3} {\bar{H}}^{\ell_1 \ell_2 \ell_3} =0~,
\eea
and
\bea
\label{feq7}
{\tilde{R}}_{ij} + \tn_{(i} h_{j)} -{1 \over 2} h_i h_j +4 Y_{i \ell_1 \ell_2} Y_j{}^{\ell_1 \ell_2}
+{1 \over 2} \Phi_{(i} {\bar{\Phi}}_{j)} -2 \xi_{(i} {\bar{\xi}}_{j)}
-{1 \over 4} H_{\ell_1 \ell_2 (i} {\bar{H}}_{j)}{}^{\ell_1 \ell_2}
\nonumber \\
+ \delta_{ij} \bigg(-{1 \over 8} \Phi_\ell {\bar{\Phi}}^\ell -{2 \over 3} Y_{\ell_1 \ell_2 \ell_3} Y^{\ell_1 \ell_2 \ell_3}
+{1 \over 48} H_{\ell_1 \ell_2 \ell_3} {\bar{H}}^{\ell_1 \ell_2 \ell_3} \bigg) =0~,
\eea
where $\tilde R$ is the Ricci tensor of ${\cal S}$.
There are three additional field equations which are not independent because they follow from those above. These are
\bea
\label{feq2}
- \tn^i L_{im} +i \Lambda^i L_{im} + h^i L_{im} -{1 \over 2} dh^{ij} H_{ijm}
+ \xi^i {\bar{L}}_{im} +{2i \over 3} (X_{m \ell_1 \ell_2 \ell_3} H^{\ell_1 \ell_3 \ell_3}
-3 Y_{m \ell_1 \ell_2} L^{\ell_1 \ell_2} ) =0~,
\nn
\eea
\bea
\label{feq6}
-{1 \over 2} \tn^j dh_{ji} - dh_{ij} h^j - \tn_i \Delta + \Delta h_i
+{4 \over 3} X_{i \ell_1 \ell_2 \ell_3} Y^{\ell_1 \ell_2 \ell_3}
\nn
-{1 \over 8} \bigg( L_{\ell_1 \ell_2} {\bar{H}}_i{}^{\ell_1 \ell_2}
-2 \Phi^\ell {\bar{L}}_{i \ell} + {\bar{L}}_{\ell_1 \ell_2} H_i{}^{\ell_1 \ell_2}
-2 {\bar{\Phi}}^\ell L_{i \ell} \bigg) =0~,
\eea
and
\bea
{1 \over 2} \tn^2 \Delta -{3 \over 2} h^i \tn_i \Delta -{1 \over 2} \Delta \hn^i h_i + \Delta h^2
+{1 \over 4} dh_{ij} dh^{ij} -{1 \over 6} X_{\ell_1 \ell_2 \ell_3 \ell_4} X^{\ell_1 \ell_2 \ell_3 \ell_4}
-{1 \over 4} L_{ij} {\bar{L}}^{ij} =0~,
\nonumber \\
\label{feq8}
\eea
which we state because they are useful in the investigation of the KSEs.

\appendix{Independent KSEs}

It is well known that the KSEs imply some of the Bianchi identities and field equations of a theory. Because of this, to find solutions it is customary to solve the
KSEs and then impose the remaining field equations and Bianchi identities. However, we shall not do this here because of the complexity of solving the KSEs  ({\ref{alg1}}), ({\ref{alg2}}), ({\ref{skse2}}), and ({\ref{alg4}}) which contain
the $\tau$ spinor as expressed in (\ref{et}). Instead, we shall first show that all the KSEs which contain $\tau_+$ are actually implied from those  containing $\phi_+$, i.e. ({\ref{skse1}}) and
({\ref{alg3}}), and some
of the field equations and Bianchi identities.

Then we also show that ({\ref{extraalg}}) and the terms linear in $u$ in ({\ref{skse1}}) and  ({\ref{alg3}})
are implied by the field equations, Bianchi identities and
({\ref{skse1x}}) and ({\ref{alg3x}}).


\subsection{ The ({\ref{skse2}}) condition} \la{(i)}

 The ({\ref{skse2}}) component of  KSEs is implied by ({\ref{skse1}}) and ({\ref{et}}) together with a number of
bosonic field equations and Bianchi identities. To see this, first evaluate the LHS of ({\ref{skse2}}) by substituting in
({\ref{et}}) to eliminate $\tau_+$, and use ({\ref{skse1}}) to evaluate the supercovariant derivatives of
$\eta_+$ and $C* \eta_+$.
Also evaluate
\bea
\label{auxelim1}
\bigg( {1 \over 4} {\tilde{R}}_{ij} \Gamma^j - {1 \over 2} \Gamma^j (\tn_j \tn_i - \tn_i \tn_j) \bigg) \phi_+
-{1 \over 2} \xi_i C* {\cal{A}}_1 \nonumber \\
-\bigg({1 \over 192} \Gamma_i{}^{\ell_1 \ell_2 \ell_3} {\bar{H}}_{\ell_1 \ell_2 \ell_3} -{3 \over 64}
{\bar{H}}_{i \ell_1 \ell_2} \Gamma^{\ell_1 \ell_2} -{1 \over 32} \Gamma_i{}^\ell {\bar{\Phi}}_\ell
+{3 \over 32}{\bar{\Phi}}_i \bigg) {\cal{A}}_1
 = 0~,
\eea
where
\bea
\label{alg3b}
{\cal{A}}_1 = \bigg(-{1 \over 4} \Phi_i \Gamma^i +{1 \over 24} H_{\ell_1 \ell_2 \ell_3} \Gamma^{\ell_1 \ell_2 \ell_3}
\bigg) \phi_+ + \xi_i \Gamma^i C* \phi_+~.
\eea
The expression in ({\ref{auxelim1}}) vanishes on making use of
({\ref{alg3}}), as ${\cal{A}}_1=0$ is equivalent to ({\ref{alg3}}).
However a non-trivial identity is obtained by expanding out the supercovariant
derivative terms again using ({\ref{skse1}}), and expanding out the ${\cal{A}}_1$ terms using
({\ref{alg3b}}).
Then, on adding ({\ref{auxelim1}}) to the LHS of ({\ref{skse2}}), with $\tau_+$ eliminated in favour of $\eta_+$
using ({\ref{et}}) and ({\ref{skse1}}) as mentioned above, one obtains, after some calculation, a term proportional to ({\ref{feq7}}).

Therefore, it follows that ({\ref{skse2}}) is implied by ({\ref{skse1}}) and ({\ref{alg3}}) and ({\ref{et}}), and the bosonic field equations and Bianchi identities.
We remark that in addition to using ({\ref{feq7}}) in establishing this identity, we also make use of
 ({\ref{ba}}), ({\ref{sd1}}), ({\ref{feq1}}) and ({\ref{feq3}}).

\subsection{The ({\ref{alg4}}) condition}
Next consider ({\ref{alg3}}) and ({\ref{alg4}}). On defining
\bea
\label{alg4b}
{\cal{A}}_2 = \bigg( -{1 \over 4} \Phi_i \Gamma^i -{1 \over 24} H_{\ell_1 \ell_2 \ell_3} \Gamma^{\ell_1 \ell_2 \ell_3} \bigg) \tau_+
- \xi_i \Gamma^i C* \tau_+ +{1 \over 8} L_{ij} \Gamma^{ij} \phi_+~,
\eea
one obtains the following identity
\bea
{\cal{A}}_2 = -{1 \over 2} \Gamma^i \tn_i {\cal{A}}_1 + \bigg( {3i \over 4} \Lambda_i \Gamma^i +{3 \over 8} h_i \Gamma^i
-{i \over 12} Y_{\ell_1 \ell_2 \ell_3} \Gamma^{\ell_1 \ell_2 \ell_3} \bigg) {\cal{A}}_1~,
\eea
where we have made use of ({\ref{skse1}}) in order to evaluate the covariant derivative in the above expression.
In addition, we also have made use of the following field equations and Bianchi identities:
({\ref{ba}}), ({\ref{feq1}}), ({\ref{feq3}}) and ({\ref{feq4}}).
It follows that these conditions, together with ({\ref{alg3}}) imply ({\ref{alg4}}).

\subsection{The ({\ref{alg1}}) condition}

To show that ({\ref{alg1}}) is also implied for the KSEs involving only $\eta$ and the field equations and Bianchi identities,  contract ({\ref{skse2}}) with $\Gamma^i$ and use ({\ref{et}})
to rewrite the $\tau_+$ terms in terms of $\phi_+$. Then subtract
$({3 \over 16}{\bar{\Phi}}_i \Gamma^i +{1 \over 96} {\bar{H}}_{\ell_1 \ell_2 \ell_3} \Gamma^{\ell_1 \ell_2 \ell_3}) {\cal{A}}_1$
from the resulting expression
to obtain ({\ref{alg1}}). In order to obtain ({\ref{alg1}}) from these expressions, we also make use of ({\ref{ba}}), ({\ref{sd1}}),
 ({\ref{feq3}}),  ({\ref{feq1}}), and ({\ref{feq5}}). It follows, from section \ref{(i)} above,  that ({\ref{alg1}})
follows from the above mentioned Bianchi identities and field equations, together with ({\ref{skse1}}) and ({\ref{alg3}}).

\subsection{The ({\ref{alg2}}) condition}

 The ({\ref{alg2}}) condition is obtained from ({\ref{alg1}}) as follows. First
act on ({\ref{alg1}}) with the Dirac operator $\Gamma^i {\tilde{\nabla}}_i$, and use the bosonic field equations and
Bianchi identities to eliminate the $d \star_8 dh$, $dL$, $d \star_8 L$, $d \star_8 h$, $dY$, $d \star_8 Y$, $dH$ and $d \star_8 H$
terms, and rewrite $d \Phi$ in terms of $L$. Then use the algebraic conditions ({\ref{alg3}}) and ({\ref{alg4}}) to eliminate the
$\xi$-terms from the resulting expression. The terms involving $\Lambda$ then vanish as a consequence of ({\ref{alg1}}).

Next consider the $dh$-terms; after some calculation, these can be rewritten
as
\bea
 {1 \over 2} dh_{ij} \Gamma^{ij} \tau_+ -{7 \over 32} h_\ell \Gamma^\ell dh_{ij} \Gamma^{ij} \phi_+
 + \big(-{1 \over 64} \Phi_\ell \Gamma^\ell +{1 \over 384} H_{\ell_1 \ell_2 \ell_3} \Gamma^{\ell_1 \ell_2 \ell_3}
 \big) dh_{ij} \Gamma^{ij} C*\phi_+ \ .
 \nonumber
 \eea
The $dh$ terms involving $\phi_+$ and $C* \phi_+$ in the above expression are then eliminated, using
({\ref{alg1}}).
On collating the remaining terms, one finds that those involving $\Delta$ (but not $d \Delta$) are
\bea
-\Delta h_j \Gamma^j \phi_+ \ .
\eea
It is also straightforward to note that the terms involving ${\bar{L}}$ vanish,
whereas the terms involving $X$ and $L$ can be rewritten as
\bea
-{1 \over 2} L_{ij} \Gamma^{ij} C* \tau_+ -{i \over 12} X_{\ell_1 \ell_2 \ell_3 \ell_4} \Gamma^{\ell_1 \ell_2 \ell_3 \ell_4} \tau_+~,
\eea
where the anti-self-duality of $X$ has been used to simplify the expression.
The remaining terms which are linear in $\tau_+, C*\tau_+$ and quadratic in $h, Y, \Phi, {\bar{\Phi}}, H, {\bar{H}}$ can be shown to vanish after some computation.
After performing these calculations, the condition which is obtained is ({\ref{alg2}}).

\subsection{The ({\ref{alg3}}) condition}
Next consider the part of ({\ref{alg3}})
which is linear in $u$. On defining
\bea
\label{alg3bx}
{\cal{B}}_1 = \bigg({1 \over 4} \Phi_i \Gamma^i +{1 \over 24} H_{\ell_1 \ell_2 \ell_3} \Gamma^{\ell_1 \ell_2 \ell_3}
\bigg) \eta_- + \xi_i \Gamma^i C* \eta_-~,
\eea
one finds that the $u$-dependent part of ({\ref{alg3}}) is proportional to
\bea
-{1 \over 2} \Gamma^i \tn_i {\cal{B}}_1 + \bigg( {3i \over 4} \Lambda_i \Gamma^i +{1 \over 8} h_i \Gamma^i
+{i \over 12} Y_{\ell_1 \ell_2 \ell_3} \Gamma^{\ell_1 \ell_2 \ell_3} \bigg) {\cal{B}}_1~,
\eea
where we have made use of ({\ref{skse1x}}) in order to evaluate the covariant derivative in the above expression.
In addition, we also have made use of the following field equations and Bianchi identities:
({\ref{ba}}),  ({\ref{feq1}}), ({\ref{feq3}}) and ({\ref{feq4}}).
It follows that these conditions, together with ({\ref{alg3x}}) imply the $u$-dependent part of ({\ref{alg3}}).

\subsection{The ({\ref{extraalg}}) condition}

Next, consider  ({\ref{extraalg}}). To show that this condition is implied
by ({\ref{skse1x}}), ({\ref{alg3x}}) and the bosonic
conditions, compute
\bea
\bigg(\Gamma^i \Gamma^j \big(\tn_j \tn_i - \tn_i \tn_j \big)
-{1 \over 2} {\tilde{R}} \bigg) \eta_-
+ \xi_i \Gamma^i C* {\cal{B}}_1
+ \bigg({5 \over 8} {\bar{\Phi}}_\ell \Gamma^\ell
-{1 \over 16} {\bar{H}}_{\ell_1 \ell_2 \ell_3}
\Gamma^{\ell_1 \ell_2 \ell_3}\bigg) {\cal{B}}_1
\eea
where we use ({\ref{skse1x}}) to evaluate the covariant derivative terms, and ({\ref{alg3x}}) implies that
the terms involving ${\cal{B}}_1$ vanish. After making
use of the field equations and Bianchi identities (specifically,  ({\ref{ba}}), ({\ref{feq1}}),
({\ref{feq3}}), ({\ref{feq5}}) and ({\ref{feq7}})), one
obtains after some calculation, a term proportional
to ({\ref{extraalg}}).

\subsection{The ({\ref{skse1}}) condition}

We next consider the part of ({\ref{skse1}}) which is
linear in $u$. First compute
\bea
\label{compare1a}
\bigg(\Gamma^j(\tn_j \tn_i - \tn_i \tn_j) \eta_-
-{1 \over 2} \Gamma^j {\tilde{R}}_{ij} \eta_-\bigg)
+ \xi_i C* {\cal{B}}_1
\nonumber \\
- \bigg(-{1 \over 16} \Gamma_i{}^q {\bar{\Phi}}_q
+{3 \over 16}{\bar{\Phi}}_i
-{1 \over 96} \Gamma_i{}^{\ell_1 \ell_2 \ell_3}
{\bar{H}}_{\ell_1 \ell_2 \ell_3}+{3 \over 32}
{\bar{H}}_{i \ell_1 \ell_2} \Gamma^{\ell_1 \ell_2}
\bigg) {\cal{B}}_1
\eea
which vanishes identically as a consequence
of ({\ref{extraalg}}). In particular, use ({\ref{skse1x}}) to evaluate
the covariant derivative terms, also using the
field equations and Bianchi identities
(specifically, ({\ref{ba}}), ({\ref{sd1}}),  ({\ref{feq1}}),
({\ref{feq3}}), ({\ref{feq5}}) and ({\ref{feq7}})).

The resulting identity obtained from
({\ref{compare1a}}) corresponds to the expression
obtained by expanding out the $u$-dependent
part of ({\ref{skse1}}), again using ({\ref{skse1x}}) to evaluate the covariant derivative terms.
Hence the $u$-dependent part of ({\ref{skse1}) is implied by
({\ref{skse1x}}) and ({\ref{extraalg}}) together with
the field equations and Bianchi identities.

\appendix{Lichnerowicz Theorems}
\setcounter{subsection}{0}

\subsection{Proof of the maximum principle on   $\parallel \eta_+\parallel^2$}

In this appendix we shall give the proof of the Lichnerowicz type theorem described in section \ref{slich1}.
Throughout the
following analysis, we  assume all of the field equations and Bianchi identities of IIB supergravity listed in appendix A.

To proceed, we rewrite the KSE ({\ref{paral}})  as
\bea
\label{kkse1}
\nabla_i^{(+)}\eta_+\equiv {\tilde{\nabla}}_i \eta_+ + \psi{}^{(+)}_i \eta_+ + \theta{}^{(+)}_i C* \eta_+ =0~,
\eea
where
\bea
\label{auxop1}
\psi{}^{(+)}_i = -{i \over 2} \Lambda_i -{1 \over 4} h_i -{i \over 4} Y_{i \ell_1 \ell_2} \Gamma^{\ell_1 \ell_2}
+{i \over 12} \Gamma_i{}^{\ell_1 \ell_2 \ell_3} Y_{\ell_1 \ell_2 \ell_3}~,
\eea
and
\bea
\label{auxop2}
\theta{}^{(+)}_i = {1 \over 16} \Gamma_i{}^j \Phi_j -{3 \over 16} \Phi_i -{1 \over 96} \Gamma_i{}^{\ell_1 \ell_2 \ell_3}
H_{\ell_1 \ell_2 \ell_3} +{3 \over 32} H_{i \ell_1 \ell_2} \Gamma^{\ell_1 \ell_2}~.
\eea
We also rewrite the associated horizon Dirac equation (\ref{dirac}) as
\bea
\label{dirac1}
{\cal D}^{(+)}\eta_+\equiv \Gamma^i {\tilde{\nabla}}_i \eta_+ + \psi{}^{(+)} \eta_+ + \theta{}^{(+)} C* \eta_+ =0~,
\eea
with
\bea
\label{auxop3}
\psi{}^{(+)} = \Gamma^i \psi{}^{(+)}_i = -{i \over 2} \Lambda_i \Gamma^i -{1 \over 4} h_i \Gamma^i +{i \over 6} Y_{\ell_1 \ell_2 \ell_3} \Gamma^{\ell_1
\ell_2 \ell_3}~,
\eea
and
\bea
\label{auxop4}
\theta{}^{(+)} = \Gamma^i \theta{}^{(+)}_i = {1 \over 4} \Phi_i \Gamma^i +{1 \over 24} H_{\ell_1 \ell_2 \ell_3} \Gamma^{\ell_1 \ell_2 \ell_3}~.
\eea

Assume that the horizon Dirac equation ({\ref{dirac1}}) holds,
we compute the Laplacian
\bea
\label{lichlap1}
\tn^i \tn_i \langle \eta_+ , \eta_+ \rangle
=2 {\rm Re} \ \bigg(\langle \eta_+, \tn^i \tn_i \eta_+ \rangle \big) + 2 \langle \tn_i \eta_+ , \tn^i \eta_+ \rangle \bigg)~.
\eea
To evaluate this expression note that
\bea
\tn^i \tn_i \eta_+ &=& \Gamma^i \tn_i \big( \Gamma^j \tn_j \eta_+ \big) +{1 \over 4} {\tilde{R}} \eta_+
\nonumber \\
&=& \Gamma^i \tn_i  \bigg(-\psi^{(+)} \eta_+ - \theta^{(+)} C* \eta_+ \bigg)
\nonumber \\
&+&{1 \over 4} \bigg(-\tn^i h_i +{1 \over 2} h^2 +{4 \over 3} Y_{\ell_1 \ell_2 \ell_3} Y^{\ell_1 \ell_2 \ell_3}
+{1 \over 2} \Phi_i {\bar{\Phi}}^i  +2 \xi_i {\bar{\xi}}^i +{1 \over 12} H_{\ell_1 \ell_2 \ell_3}
{\bar{H}}^{\ell_1 \ell_2 \ell_3} \bigg) \langle \eta_+ , \eta_+ \rangle~.
\nonumber \\
\eea
It follows that
\bea
{\rm Re} \ \big( \langle \eta_+ , \tn^i \tn_i \eta_+ \rangle \big)
&=& {1 \over 4} \bigg(-\tn^i h_i +{1 \over 2} h^2 +{4 \over 3} Y_{\ell_1 \ell_2 \ell_3} Y^{\ell_1 \ell_2 \ell_3}
\nonumber \\
&+&{1 \over 2} \Phi_i {\bar{\Phi}}^i  +2 \xi_i {\bar{\xi}}^i +{1 \over 12} H_{\ell_1 \ell_2 \ell_3}
{\bar{H}}^{\ell_1 \ell_2 \ell_3} \bigg) \langle \eta_+ , \eta_+ \rangle
\nonumber \\
&+& \langle \eta_+ , \bigg({1 \over 4} \tn^i h_i -{i \over 2} \tn^i Y_{i \ell_1 \ell_2} \Gamma^{\ell_1 \ell_2}
+{i \over 4} (d \Lambda)_{ij} \Gamma^{ij} \bigg) \eta_+ \rangle
\nonumber \\
&+& {\rm Re} \ \langle \eta_+, \bigg(-{1 \over 4} \tn^i \Phi_i -{1 \over 96} (dH)_{\ell_1 \ell_2 \ell_3 \ell_4}
\Gamma^{\ell_1 \ell_2 \ell_3 \ell_4} \bigg) C* \eta_+ \rangle
\nonumber \\
&+& {\rm Re} \ \langle \eta_+, \Gamma^i \bigg({i \over 2} \Lambda_j \Gamma^j +{1 \over 4} h_j \Gamma^j
-{i \over 6} Y_{\ell_1 \ell_2 \ell_3} \Gamma^{\ell_1 \ell_2 \ell_3} \bigg) \tn_i \eta_+ \rangle
\nonumber \\
&+& {\rm Re} \ \langle \eta_+, \Gamma^i \bigg(-{1 \over 4} \Phi_j \Gamma^j -{1 \over 24} H_{\ell_1 \ell_2 \ell_3}
\Gamma^{\ell_1 \ell_2 \ell_3} \bigg) C* \tn_i \eta_+ \rangle~.
\eea
Using the field equations and Bianchi identities, the terms involving $\tn^i Y_{i \ell_1 \ell_2}, d \Lambda, \tn^i \Phi_i, dH$
can be rewritten as terms quadratic in fluxes to give
\bea
\label{lichaux1}
{\rm Re} \ \big( \langle \eta_+ , \tn^i \tn_i \eta_+ \rangle \big)
&=& \bigg({1 \over 8} h^2 +{1 \over 3} Y_{\ell_1 \ell_2 \ell_3} Y^{\ell_1 \ell_2 \ell_3}
+{1 \over 8} \Phi_i {\bar{\Phi}}^i +{1 \over 2} \xi_i {\bar{\xi}}^i +{1 \over 48}
H_{\ell_1 \ell_2 \ell_3} {\bar{H}}^{\ell_1 \ell_2 \ell_3} \bigg) \langle \eta_+ , \eta_+ \rangle
\nonumber \\
&+& \langle \eta_+ , \bigg( {1 \over 288} H_{\ell_1 \ell_2 \ell_3} {\bar{H}}_{\ell_4 \ell_5 \ell_6}
\Gamma^{\ell_1 \ell_2 \ell_3 \ell_4 \ell_5 \ell_6}+{1 \over 2} \xi_{\ell_1} {\bar{\xi}}_{\ell_2}
\Gamma^{\ell_1 \ell_2} \bigg) \eta_+ \rangle
\nonumber \\
&+& {\rm Re} \ \langle \eta_+ , \bigg(-{i \over 4} \Lambda^i \Phi_i
-{1 \over 4} \xi^i {\bar{\Phi}}_i +{i \over 6} Y_{\ell_1 \ell_2 \ell_3} H^{\ell_1 \ell_2 \ell_3}
\nonumber \\
&-&{i \over 24} \Lambda_{\ell_1} H_{\ell_1 \ell_3 \ell_4} \Gamma^{\ell_1 \ell_2 \ell_3 \ell_4}
+{1 \over 24} \xi_{\ell_1} {\bar{H}}_{\ell_2 \ell_3 \ell_4} \Gamma^{\ell_1 \ell_2 \ell_3 \ell_4}
\bigg) C* \eta_+ \rangle
\nonumber \\
&+& {\rm Re} \ \langle \eta_+, \Gamma^i \bigg({i \over 2} \Lambda_j \Gamma^j +{1 \over 4} h_j \Gamma^j
-{i \over 6} Y_{\ell_1 \ell_2 \ell_3} \Gamma^{\ell_1 \ell_2 \ell_3} \bigg) \tn_i \eta_+ \rangle
\nonumber \\
&+& {\rm Re} \ \langle \eta_+, \Gamma^i \bigg(-{1 \over 4} \Phi_j \Gamma^j -{1 \over 24} H_{\ell_1 \ell_2 \ell_3}
\Gamma^{\ell_1 \ell_2 \ell_3} \bigg) C* \tn_i \eta_+ \rangle~,
\eea
where we have made use of the identities
\bea
\langle \eta_+, \Gamma_{ij} C* \eta_+ \rangle =0 \ .
\eea
Having evaluated the first term in ({\ref{lichlap1}}), we compute the second term, writing
\bea
\label{lichaux2}
\langle \tn^i \eta_+ , \tn_i \eta_+ \rangle &=&
\langle \tn^i \eta_+ + \psi{}^{(+)}{}^i \eta_+ + \theta{}^{(+)}{}^i C* \eta_+, \tn_i \eta_+ + \psi{}^{(+)}_i \eta_+
+ \theta{}^{(+)}_i C* \eta_+ \rangle
\nonumber \\
&-& \langle \psi{}^{(+)}{}^i \eta_+ + \theta{}^{(+)}{}^i C* \eta_+, \psi{}^{(+)}_i \eta_+ + \theta{}^{(+)}_i C* \eta_+ \rangle
\nonumber \\
&+& {\rm Re} \ \langle \eta_+, -2(\psi{}^{(+)}{}^i){}^\dagger \tn_i \eta_+ -2 (C* (\theta{}^{(+)}{}^i){}^\dagger) \tn_i C* \eta_+ \rangle~.
\eea
The first term on the RHS is the norm squared of the horizon KSE ({\ref{skse1}}).
Next, combine the terms which involve $\tn_i \eta_+$ and $\tn_i C* \eta_+$ in the last two lines of ({\ref{lichaux1}})
and the last line of ({\ref{lichaux2}}). These can be rewritten using the identities
\bea
\Gamma^i({i \over 2} \Lambda_j \Gamma^j +{1 \over 4}h^j \Gamma^j
-{i \over 6} Y_{\ell_1 \ell_2 \ell_3} \Gamma^{\ell_1 \ell_2 \ell_3}) -2 (\psi{}^{(+)}{}^i){}^\dagger
= h^i + \bigg(-{i \over 2} \Lambda^j \Gamma^j -{1 \over 4} h_j \Gamma^j \bigg) \Gamma^i~,
\eea
and
\bea
\Gamma^i (-{1 \over 4} \Phi_j \Gamma^j -{1 \over 24} H_{\ell_1 \ell_2 \ell_3}
\Gamma^{\ell_1 \ell_2 \ell_3}) -2 (C* (\theta{}^{(+)}{}^i){}^\dagger)
= \bigg({1 \over 8} \Phi_j \Gamma^j +{1 \over 48} H_{\ell_1 \ell_2 \ell_3}
\Gamma^{\ell_1 \ell_2 \ell_3} \bigg) \Gamma^i~.
\eea
It follows that the sum of the last two lines in ({\ref{lichaux1}}) and the last line in ({\ref{lichaux2}})
can be rewritten, using the horizon Dirac equation, to give
\bea
\label{lichaux3}
&& {\rm Re} \ \langle \eta_+, \Gamma^i \bigg({i \over 2} \Lambda_j \Gamma^j +{1 \over 4} h_j \Gamma^j
-{i \over 6} Y_{\ell_1 \ell_2 \ell_3} \Gamma^{\ell_1 \ell_2 \ell_3} \bigg) \tn_i \eta_+ \rangle
\nonumber \\
&+& {\rm Re} \ \langle \eta_+, \Gamma^i \bigg(-{1 \over 4} \Phi_j \Gamma^j -{1 \over 24} H_{\ell_1 \ell_2 \ell_3}
\Gamma^{\ell_1 \ell_2 \ell_3} \bigg) C* \tn_i \eta_+ \rangle
\nonumber \\
&+& {\rm Re} \ \langle \eta_+, -2(\psi{}^{(+)}{}^i){}^\dagger \tn_i \eta_+ -2 (C* (\theta{}^{(+)}{}^i){}^\dagger) \tn_i C* \eta_+ \rangle
\nonumber \\
&=& {1 \over 2} h^i \tn_i \langle \eta_+ , \eta_+ \rangle
\nonumber \\
&+& {\rm Re} \ \langle \eta_+ ,  \bigg(-{i \over 2} \Lambda^j \Gamma^j -{1 \over 4} h_j \Gamma^j \bigg)
\bigg(-\psi{}^{(+)} \eta_+ - \theta{}^{(+)} C* \eta_+ \bigg) \rangle
\nonumber \\
&+& {\rm Re} \ \langle \eta_+ , \bigg({1 \over 8} \Phi_j \Gamma^j +{1 \over 48} H_{\ell_1 \ell_2 \ell_3}
\Gamma^{\ell_1 \ell_2 \ell_3} \bigg)  \bigg(-(C* \psi{}^{(+)}) C* \eta_+ - (C* \theta{}^{(+)}) \eta_+ \bigg) \rangle~.
\nonumber \\
\eea

So, to evaluate ({\ref{lichlap1}}) one takes the sum of ({\ref{lichaux1}}) and ({\ref{lichaux2}}),
using ({\ref{lichaux3}}) to rewrite the last two lines in ({\ref{lichaux1}}) and the last line in ({\ref{lichaux2}})
as given above. On expanding out all of the terms quadratic in the fluxes, one obtains (\ref{lichident}).

\subsection{Proof of a Lichnerowicz Theorem for ${\cal D}^{(-)}$}

In this appendix,  we shall prove a  Lichnerowicz identity for the ${\cal D}^{(-)}$ operator in (\ref{dirac}).  We remark that throughout the
following analysis, we will again assume all of the Bianchi identities and field equations which are listed in appendix A.

To proceed, we rewrite the KSE ({\ref{paral}}) as
\bea
\label{kkse1x}
\nabla^{(-)}\eta_-\equiv {\tilde{\nabla}}_i \eta_- + \psi{}^{(-)}_i \eta_- + \theta{}^{(-)}_i C* \eta_-~,
\eea
where
\bea
\label{auxop1x}
\psi{}^{(-)}_i = -{i \over 2} \Lambda_i +{1 \over 4} h_i +{i \over 4} Y_{i \ell_1 \ell_2} \Gamma^{\ell_1 \ell_2}
-{i \over 12} \Gamma_i{}^{\ell_1 \ell_2 \ell_3} Y_{\ell_1 \ell_2 \ell_3}~,
\eea
and
\bea
\label{auxop2x}
\theta{}^{(-)}_i = -{1 \over 16} \Gamma_i{}^j \Phi_j +{3 \over 16} \Phi_i -{1 \over 96} \Gamma_i{}^{\ell_1 \ell_2 \ell_3}
H_{\ell_1 \ell_2 \ell_3} +{3 \over 32} H_{i \ell_1 \ell_2} \Gamma^{\ell_1 \ell_2}~.
\eea
We also rewrite the associated horizon Dirac equation (\ref{dirac})  as
\bea
\label{dirac1x}
{\cal D}^{(-)}\eta_-\equiv \Gamma^i {\tilde{\nabla}}_i \eta_- + \psi{}^{(-)} \eta_- + \theta{}^{(-)} C* \eta_-~,
\eea
with
\bea
\label{auxop3x}
\psi{}^{(-)} = \Gamma^i \psi{}^{(-)}_i = -{i \over 2} \Lambda_i \Gamma^i +{1 \over 4} h_i \Gamma^i -{i \over 6} Y_{\ell_1 \ell_2 \ell_3} \Gamma^{\ell_1
\ell_2 \ell_3}~,
\eea
and
\bea
\label{auxop4x}
\theta{}^{(-)} = \Gamma^i \theta{}^{(-)}_i = -{1 \over 4} \Phi_i \Gamma^i +{1 \over 24} H_{\ell_1 \ell_2 \ell_3} \Gamma^{\ell_1 \ell_2 \ell_3}~.
\eea

We define
\bea
\label{lichn1}
{\cal{I}} &=& \int_{{\cal{S}}} \big ( \parallel \nabla^{(-)}\eta_-\parallel^2-\parallel{\cal D}^{(-)}\eta_-\parallel^2\big)~,
 \eea
 and decompose
\bea
\label{lichner1}
{\cal{I}}= {\cal{I}}_1 + {\cal{I}}_2 + {\cal{I}}_3~,
\eea
where
\bea
\label{termx1}
{\cal{I}}_1 = \int_{{\cal{S}}} \langle {\tilde{\nabla}}_i \eta_- ,
{\tilde{\nabla}}^i \eta_- \rangle
- \langle \Gamma^i {\tilde{\nabla}}_i \eta_- ,
 \Gamma^j {\tilde{\nabla}}_j \eta_- \rangle~.
 \eea
 and
 \bea
 \label{termx2}
 {\cal{I}}_2  = 2 {\rm Re} \ \bigg( \int_{{\cal{S}}} \langle {\tilde{\nabla}}_i \eta_-, \Psi{}^{(-)}{}^i \eta_-  \rangle
 - \langle \Gamma^i {\tilde{\nabla}}_i \eta_- , \Psi{}^{(-)} \eta_-  \rangle \bigg)~,
\eea
and
\bea
\label{termx3}
{\cal{I}}_3 = \int_{{\cal{S}}} \langle  \Psi{}^{(-)}_i \eta_-  ,
 \Psi{}^{(-)}{}^i \eta_-  \rangle
- \langle  \Psi{}^{(-)} \eta_-  , \Psi{}^{(-)} \eta_- \rangle~.
\nonumber \\
\eea
On using the identity
\bea
\langle C* \eta, \tau \rangle = {\overline{ \langle \eta, C* \tau \rangle }}~,
\eea
for any $\eta, \tau$,
it is straightforward to rewrite ({\ref{termx3}) as
\bea
\label{termx3b}
{\cal{I}}_3 &=& \int_{\cal{S}} \langle \eta_- , \bigg( (\psi{}^{(-)}_i)^\dagger \psi{}^{(-)}{}^i - \psi{}^{(-)}{}^\dagger \psi{}^{(-)}
+ C* \big( (\theta{}^{(-)}{}^i)^\dagger \theta{}^{(-)}_i - \theta{}^{(-)}{}^\dagger \theta{}^{(-)}  \big) \bigg) \eta_- \rangle
\nonumber \\
&+& 2 {\rm Re} \ \int_{\cal{S}} \langle \eta_- , \bigg( (\psi{}^{(-)}_i)^\dagger \theta{}^{(-)}{}^i - \psi{}^{(-)}{}^\dagger \theta{}^{(-)} \bigg) C* \eta_- \rangle~,
\eea
where in the first line of the above expression, the charge conjugation acts solely on $(\theta{}^{(-)}{}^i)^\dagger \theta{}^{(-)}_i - \theta{}^{(-)}{}^\dagger \theta{}^{(-)}$.

To proceed we expand out ({\ref{termx3b}}) to obtain
\bea
\label{termx3c}
{\cal{I}}_3 &=& \int_{\cal{S}} \langle \eta_- , \bigg({1 \over 6} Y_{\ell_1 \ell_2 \ell_3} Y^{\ell_1 \ell_2 \ell_3}
+{1 \over 96} H_{\ell_1 \ell_2 \ell_3} {\bar{H}}^{\ell_1 \ell_2 \ell_3}
+ \big( -{i \over 4} \Lambda_{\ell_1} h_{\ell_2}
+{3i \over 8} h^i Y_{i \ell_1 \ell_2}
\nonumber \\
&-&{1 \over 16} \Phi_{\ell_1} {\bar{\Phi}}_{\ell_2}
+{3 \over 64} \Phi^i {\bar{H}}_{i \ell_1 \ell_2} -{3 \over 64} {\bar{\Phi}}^i H_{i \ell_1 \ell_2}
\nonumber \\
&+&{1 \over 576} \epsilon_{\ell_1 \ell_2}{}^{q_1 q_2 q_3 q_4 q_5 q_6}
H_{q_1 q_2 q_3} {\bar{H}}_{q_4 q_5 q_6} \big) \Gamma^{\ell_1 \ell_2}
\nonumber \\
&+& \big( -{1 \over 12} \Lambda_{\ell_1} Y_{\ell_2 \ell_3 \ell_4}
+{1 \over 4} Y^i{}_{\ell_1 \ell_2} Y_{i \ell_3 \ell_4}
+{1 \over 192} \Phi_{\ell_1} {\bar{H}}_{\ell_2 \ell_3 \ell_4}
\nonumber \\
&+&{1 \over 192} {\bar{\Phi}}_{\ell_1} H_{\ell_2 \ell_3 \ell_4}
+{1 \over 64} H_{i \ell_1 \ell_2} {\bar{H}}^i{}_{\ell_3 \ell_4} \big) \Gamma^{\ell_1 \ell_2 \ell_3 \ell_4}  \bigg) \eta_- \rangle
\nonumber \\
&+& 2 {\rm Re} \ \int_{\cal{S}} \langle \eta_- , \bigg( {7i \over 32} \Lambda^i \Phi_i +{7 \over 64} h^i \Phi_i
-{11i \over 96} Y_{\ell_1 \ell_2 \ell_3} H^{\ell_1 \ell_2 \ell_3}
\nonumber \\
&+& \big( -{5i \over 192} \Lambda_{\ell_1} H_{\ell_2 \ell_3 \ell_4}
-{5 \over 384} h_{\ell_1} H_{\ell_2 \ell_3 \ell_4}
\nonumber \\
&+&{5i \over 96} \Phi_{\ell_1} Y_{\ell_2 \ell_3 \ell_4}
+{3i \over 64} Y_{i \ell_1 \ell_2} H^i{}_{\ell_3 \ell_4} \big) \Gamma^{\ell_1 \ell_2 \ell_3 \ell_4} \bigg) C* \eta_- \rangle~,
\eea
where we have made use of the identities
\bea
\langle \eta_- , \Gamma_{\ell_1 \ell_2} C* \eta_- \rangle =0 , \qquad \langle \eta_- , \Gamma_{\ell_1 \ell_2 \ell_3 \ell_4 \ell_5 \ell_6} C* \eta_- \rangle =0~.
\eea

It is also straightforward to evaluate ${\cal{I}}_1$, to obtain
\bea
\label{termx2c}
{\cal{I}}_1 &=& \int_{\cal{S}} - {\tilde{\nabla}}_i \langle \eta_-, \Gamma^{ij} {\tilde{\nabla}}_j \eta_- \rangle
-{1 \over 4} \int_{\cal{S}} h^i {\tilde{\nabla}}_i \langle \eta_- , \eta_- \rangle
\nonumber \\
&+& \int_{\cal{S}} \langle \eta_- , \big( -{1 \over 8} h^2 -{1 \over 3} Y_{\ell_1 \ell_2 \ell_3} Y^{\ell_1 \ell_2 \ell_3}
-{1 \over 8} \Phi_i {\bar{\Phi}}^i
\nonumber \\
 &-&{1 \over 2} \xi_i {\bar{\xi}}^i -{1 \over 48} H_{\ell_1 \ell_2 \ell_3} {\bar{H}}^{\ell_1 \ell_2 \ell_3}
\big) \eta_- \rangle~,
\eea
where we have used the Einstein equations ({\ref{feq7}}) to compute
\bea
{\tilde{R}}  = -{\tilde{\nabla}}^i h_i + {1 \over 2} h^2 +{4 \over 3} Y_{\ell_1 \ell_2 \ell_3} Y^{\ell_1 \ell_2 \ell_3}
+{1 \over 2} \Phi_i {\bar{\Phi}}^i +2 \xi_i {\bar{\xi}}^i +{1 \over 12} H_{\ell_1 \ell_2 \ell_3} {\bar{H}}^{\ell_1 \ell_2 \ell_3}~,
\eea
and we recall
\bea
\Gamma^{ij} {\tilde{\nabla}}_i {\tilde{\nabla}}_j \eta_- = -{1 \over 4} {\tilde{R}} \eta_- \ .
\eea

It remains to compute ${\cal{I}}_2$.  First note that

\bea
\label{ttr1}
{\cal{I}}_2 &=& \int_{{\cal{S}}} \tn_i \langle \eta_-, {i \over 2} \Gamma^{ij} \Lambda_j \eta_- \rangle
\nonumber \\
&+& \int_{\cal{S}} \langle \eta_-, \big(-\tn_i \psi{}^{(-)}{}^i + \tn_i (\Gamma^i \psi{}^{(-)}) \big) \eta_- \rangle
\nonumber \\
&+& \int_{\cal{S}} \langle \eta_- , \big( (\psi{}^{(-)}{}^i)^\dagger - \psi{}^{(-)}{}^i - (\psi{}^{(-)}{}^\dagger -\psi{}^{(-)}) \Gamma^i \big) \tn_i \eta_- \rangle
\nonumber \\
&+& \int_{\cal{S}} \langle \eta_- , \big( \Gamma^i \psi{}^{(-)} - \psi{}^{(-)} \Gamma^i \big) \tn_i \eta_- \rangle
\nonumber \\
&+& {\rm Re \ } \int_{\cal{S}} \langle \eta_- , \big(-2 \tn_i \theta{}^{(-)}{}^i + 2 \tn_i (\Gamma^i \theta{}^{(-)}) \big) C* \eta_-
+ C* \bigg( \big( (\tn_i \theta{}^{(-)}{}^i)^\dagger - \tn_i (\Gamma^i \theta{}^{(-)})^\dagger \big) \eta_- \bigg) \rangle
\nonumber \\
&+& {\rm Re \ } \int_{\cal{S}} \langle \eta_- , \big( -\theta{}^{(-)}{}^i + \Gamma^i \theta{}^{(-)} \big) \tn_i C* \eta_-
+ C* \bigg( \big((\theta{}^{(-)}{}^i)^\dagger -(\Gamma^i \theta{}^{(-)})^\dagger \big) \tn_i \eta_- \bigg) \rangle~.
\eea

In order to evaluate this expression, it is useful to note that
\bea
\label{ttr1a}
{\rm Re \ } \big( \int_{\cal{S}} \langle \eta_-, \big(-\tn_i \psi{}^{(-)}{}^i + \tn_i (\Gamma^i \psi{}^{(-)}) \big) \eta_- \rangle \big)
= \int_{\cal{S}} \langle \eta_- , (-{i \over 4} d \Lambda_{\ell_1 \ell_2} \Gamma^{\ell_1 \ell_2}
-{3i \over 4} \tn^i Y_{i \ell_1 \ell_2} \Gamma^{\ell_1 \ell_2} \big) \eta_- \rangle~,
\nonumber \\
\eea
where, as a consequence of ({\ref{ba}}),
\bea
\langle \eta_- , d \Lambda_{\ell_1 \ell_2} \Gamma^{\ell_1 \ell_2} \eta_- \rangle =
-2i \langle {\bar{\xi}}_i \Gamma^i \eta_-, {\bar{\xi}}_j \Gamma^j \eta_- \rangle +2i \langle \eta_-, \xi^i {\bar{\xi}}_i \eta_- \rangle~.
\eea

Also, one has
\bea
\label{ttr1b}
 \int_{\cal{S}} \langle \eta_- , \big( (\psi{}^{(-)}{}^i)^\dagger - \psi{}^{(-)}{}^i - (\psi{}^{(-)}{}^\dagger -\psi{}^{(-)}) \Gamma^i \big) \tn_i \eta_- \rangle
+ \int_{\cal{S}} \langle \eta_- , \big( \Gamma^i \psi{}^{(-)} - \psi{}^{(-)} \Gamma^i \big) \tn_i \eta_- \rangle
\nonumber \\
= \int_{\cal{S}} \langle \eta_- , \big( (-{1 \over 2} h_j \Gamma^j +{i \over 6} Y_{\ell_1 \ell_2 \ell_3}
\Gamma^{\ell_1 \ell_2 \ell_3}) \Gamma^i +{1 \over 2} h^i -{i \over 2} Y^i{}_{\ell_1 \ell_2} \big) \tn_i \eta_- \rangle~,
\nonumber \\
\eea
with
\bea
{\rm Re \ } \int_{\cal{S}} \langle \eta_- , h^i \tn_i \eta_- \rangle = {1 \over 2} \int_{\cal{S}} h^i \tn_i \langle \eta_-, \eta_- \rangle~,
\eea
and
\bea
{\rm Re \ } \int_{\cal{S}} \langle \eta_- , i Y^i{}_{\ell_1 \ell_2} \Gamma^{\ell_1 \ell_2} \tn_i \eta_- \rangle
=-{1 \over 2} \int_{\cal{S}} \langle \eta_-, i \tn^i Y_{i \ell_1 \ell_2} \Gamma^{\ell_1 \ell_2} \eta_- \rangle~,
\eea
and hence ({\ref{ttr1b}}) implies
\bea
\label{ttr1c}
{\rm Re \ } \bigg(\int_{\cal{S}} \langle \eta_- , \big( (\psi{}^{(-)}{}^i)^\dagger - \psi{}^{(-)}{}^i - (\psi{}^{(-)}{}^\dagger -\psi{}^{(-)}) \Gamma^i \big) \tn_i \eta_- \rangle
+ \int_{\cal{S}} \langle \eta_- , \big( \Gamma^i \psi{}^{(-)} - \psi{}^{(-)} \Gamma^i \big) \tn_i \eta_- \rangle \bigg)
\nonumber \\
={\rm Re} \int_{\cal{S}} \langle \eta_-, (-{1 \over 2} h_j \Gamma^j +{i \over 6} Y_{\ell_1 \ell_2 \ell_3} \Gamma^{\ell_1 \ell_2 \ell_3}
) \Gamma^i \tn_i \eta_- \rangle
+{1 \over 4} \int_{\cal{S}} h^i \tn_i \langle \eta_- , \eta_- \rangle
\nonumber \\
+{1 \over 4} \int_{\cal{S}} \langle \eta_- , i \tn^i Y_{i \ell_1 \ell_2} \Gamma^{\ell_1 \ell_2} \eta_- \rangle~.
\nonumber \\
\eea
In addition, one has
\bea
\label{ttr1d}
 {\rm Re \ } \int_{\cal{S}} \langle \eta_- , \big(-2 \tn_i \theta{}^{(-)}{}^i + 2 \tn_i (\Gamma^i \theta{}^{(-)}) \big) C* \eta_-
+ C* \bigg( \big( (\tn_i \theta{}^{(-)}{}^i)^\dagger - \tn_i (\Gamma^i \theta{}^{(-)})^\dagger \big) \eta_- \bigg) \rangle
\nonumber \\
= {\rm Re \ } \int_{\cal{S}} \langle \eta_-, \big( -{7 \over 16} \tn^i \Phi_i +{5 \over 384} dH_{\ell_1 \ell_2 \ell_3 \ell_4} \Gamma^{\ell_1 \ell_2 \ell_3 \ell_4}
\big) C* \eta_- \rangle~.
\nonumber \\
\eea

Also,
\bea
\label{ttr1e}
 {\rm Re \ } \int_{\cal{S}} \langle \eta_- , \big( -\theta{}^{(-)}{}^i + \Gamma^i \theta{}^{(-)} \big) \tn_i C* \eta_-
+ C* \bigg( \big((\theta{}^{(-)}{}^i)^\dagger -(\Gamma^i \theta{}^{(-)})^\dagger \big) \tn_i \eta_- \bigg) \rangle
\nonumber \\
= {\rm Re \ } \int_{\cal{S}} \langle \eta_- , \big({3 \over 8} \Gamma^j  \Phi_j
+{1 \over 48} H_{\ell_1 \ell_2 \ell_3} \Gamma^{\ell_1 \ell_2 \ell_3} \big) \Gamma^i \tn_i C* \eta_-
\nonumber \\
-{3 \over 8} \Phi^i \tn_i C* \eta_- +{1 \over 48} \Gamma^{i \ell_1 \ell_2 \ell_3} H_{\ell_1 \ell_2 \ell_3} \tn_i
C*\eta_- \rangle~,
\eea
where
\bea
\label{auxtt1}
{\rm Re \ } \int_{\cal{S}} \langle \eta_- , \Phi^i \tn_i C* \eta_- \rangle = -{1 \over 2} {\rm Re \ } \int_{\cal{S}}
\langle  \eta_-, (\tn^i \Phi_i) C* \eta_- \rangle~,
\eea
and
\bea
\label{auxtt2}
{\rm Re \ } \int_{\cal{S}} \langle \eta_-, \Gamma^{i \ell_1 \ell_2 \ell_3} H_{\ell_1 \ell_2 \ell_3} \tn_i C* \eta_- \rangle
= -{1 \over 8} {\rm Re \ } \int_{\cal{S}} \langle \eta_- , dH_{\ell_1 \ell_2 \ell_3 \ell_4} \Gamma^{\ell_1 \ell_2 \ell_3 \ell_4}
 C* \eta_- \rangle~,
 \eea
 and we remark that the surface terms obtained on integrating by parts in ({\ref{auxtt1}}) and ({\ref{auxtt2}})
 vanish.

 It follows that ({\ref{ttr1e}}) can be rewritten as
 \bea
 \label{ttr1f}
 {\rm Re \ } \int_{\cal{S}} \langle \eta_- , \big( -\theta{}^{(-)}{}^i + \Gamma^i \theta{}^{(-)} \big) \tn_i C* \eta_-
+ C* \bigg( \big((\theta{}^{(-)}{}^i)^\dagger -(\Gamma^i \theta{}^{(-)})^\dagger \big) \tn_i \eta_- \bigg) \rangle
\nonumber \\
= {\rm Re \ } \int_{\cal{S}} \langle \eta_- , \big({3 \over 8} \Gamma^j  \Phi_j
+{1 \over 48} H_{\ell_1 \ell_2 \ell_3} \Gamma^{\ell_1 \ell_2 \ell_3} \big) \Gamma^i \tn_i C* \eta_- \rangle
\nonumber \\
+ {\rm Re \ } \int_{\cal{S}} \langle \eta_-, \big({3 \over 16} \tn^i \Phi_i -{1 \over 384} dH_{\ell_1 \ell_2 \ell_3 \ell_4}
\Gamma^{\ell_1 \ell_2 \ell_3 \ell_4} \big) C* \eta_- \rangle~.
\eea

 On substituting ({\ref{ttr1a}}), ({\ref{ttr1c}}), ({\ref{ttr1d}}) and ({\ref{ttr1f}}) into ({\ref{ttr1}}), one obtains,
 after using the bosonic Bianchi identities/field equations,

\bea
\label{ttr2}
{\cal{I}}_2 &=& \int_{\cal{S}} \tn_i \langle \eta_-, {i \over 2} \Gamma^{ij} \Lambda_j \eta_- \rangle
+{1 \over 2} \langle \eta_-, \xi_i {\bar{\xi}}^i  \eta_- \rangle
\nonumber \\
&+& \int_{\cal{S}} -{1 \over 2} \langle {\bar{\xi}}_i \Gamma^i \eta_-, {\bar{\xi}}_j \Gamma^j \eta_- \rangle
+{1 \over 4} \int_{\cal{S}} h^i \tn_i \langle \eta_-, \eta_- \rangle
\nonumber \\
&+& \int_{\cal{S}} \langle \eta_- , -{1 \over 576} \epsilon_{\ell_1 \ell_2}{}^{q_1 q_2 q_3 q_4 q_5 q_6}
H_{q_1 q_2 q_3} {\bar{H}}_{q_4 q_5 q_6} \Gamma^{\ell_1 \ell_2} \eta_- \rangle
\nonumber \\
&+& {\rm Re \ } \int_{\cal{S}} \langle \eta_-, \big(-{1 \over 2} h_j \Gamma^j +{i \over 6} Y_{\ell_1 \ell_2 \ell_3}
\Gamma^{\ell_1 \ell_2 \ell_3} \big) \Gamma^i \tn_i \eta_- \rangle
\nonumber \\
&+& {\rm Re \ } \int_{\cal{S}} \langle \eta_- , \big({3 \over 8} \Phi_j \Gamma^j +{1 \over 48} H_{\ell_1 \ell_2 \ell_3}
\Gamma^{\ell_1 \ell_2 \ell_3} \big) \Gamma^i \tn_i C* \eta_- \rangle
\nonumber \\
&+& {\rm Re \ } \int_{\cal{S}} \langle \eta_- , {1 \over 4} \big( -i \Lambda^i \Phi_i - \xi^i {\bar{\Phi}}_i
+{2i \over 3} H_{\ell_1 \ell_2 \ell_3} Y^{\ell_1 \ell_2 \ell_3} \big) C* \eta_- \rangle
\nonumber \\
&+& {\rm Re \ } \int_{\cal{S}} \langle \eta_- , {1 \over 96} \big( 4i \Lambda_{\ell_1} H_{\ell_2 \ell_3 \ell_4}
-4 \xi_{\ell_1} {\bar{H}}_{\ell_2 \ell_3 \ell_4} \big) \Gamma^{\ell_1 \ell_2 \ell_3 \ell_4} C* \eta_- \rangle~.
\eea
To proceed further, complete the square involving the $\xi$-terms in lines 2, 6 and 7 of the above expression.
This produces a term proportional to the norm squared of the LHS of the algebraic condition ({\ref{alg3x}}),
together with a number of counterterms. Also rewrite lines 4 and 5 in terms of the horizon Dirac equation
({\ref{dirac1x}}) and its conjugate (with respect to $C*$); again there are a number of algebraic counterterms.
On performing these calculations, one obtains
\bea
\label{ttr3}
{\cal{I}}_2 &=& \int_{\cal{S}} \tn_i \langle \eta_-, {i \over 2} \Gamma^{ij} \Lambda_j \eta_- \rangle
+{1 \over 4} \int_{\cal{S}} h^i \tn_i \langle \eta_-, \eta_- \rangle
\nonumber \\
&+& \int_{\cal{S}} -{1 \over 2} \langle {\bar{\xi}}_i \Gamma^i \eta_- + ({1 \over 4} {\bar{\Phi}}_i \Gamma^i
+{1 \over 24} {\bar{H}}_{\ell_1 \ell_2 \ell_3} \Gamma^{\ell_1 \ell_2 \ell_3} ) C* \eta_- ,
\nonumber \\
& & \ \ \ \ \ \ \ \ \
{\bar{\xi}}_j \Gamma^j \eta_- + ({1 \over 4} {\bar{\Phi}}_j \Gamma^j
+{1 \over 24} {\bar{H}}_{q_1 q_2 q_3} \Gamma^{q_1 q_2 q_3} ) C* \eta_- \rangle
\nonumber \\
&+& {\rm Re \ } \int_{\cal{S}} \langle \eta_- , \big(-{1 \over 2} h_j \Gamma^j +{i \over 6} Y_{\ell_1 \ell_2 \ell_3}
\Gamma^{\ell_1 \ell_2 \ell_3} \big) \big( \Gamma^i \tn_i \eta_- + \psi{}^{(-)} \eta_- + \theta{}^{(-)} C* \eta_- \big) \rangle
\nonumber \\
&+& {\rm Re \ } \int_{\cal{S}} \langle \eta_- , \big({3 \over 8} \Phi_j \Gamma^j +{1 \over 48} H_{\ell_1 \ell_2 \ell_3}
\Gamma^{\ell_1 \ell_2 \ell_3} \big) C* \big( \Gamma^i \tn_i \eta_- + \psi{}^{(-)} \eta_- + \theta{}^{(-)} C* \eta_- \big) \rangle
\nonumber \\
&+& \int_{\cal{S}} \langle \eta_- , \bigg( {1 \over 2} \xi_i {\bar{\xi}}_i +{1 \over 8} h^2
+{1 \over 6} Y_{\ell_1 \ell_2 \ell_3} Y^{\ell_1 \ell_2 \ell_3}
+{1 \over 8} \Phi_i {\bar{\Phi}}^i +{1 \over 96} H_{\ell_1 \ell_2 \ell_3} {\bar{H}}^{\ell_1 \ell_2 \ell_3}
\nonumber \\
&+& \big({i \over 4} \Lambda_{\ell_1} h_{\ell_2}
+{3i \over 8} h^i Y_{i \ell_1 \ell_2}
+{1 \over 16} \Phi_{\ell_1} {\bar{\Phi}}_{\ell_2}
-{3 \over 64} \Phi^i {\bar{H}}_{i \ell_1 \ell_2}
+{3 \over 64} {\bar{\Phi}}^i H_{i \ell_1 \ell_2}
\nonumber \\
&-&{1 \over 576} \epsilon_{\ell_1 \ell_2}{}^{q_1 q_2 q_3 q_4 q_5 q_6} H_{q_1 q_2 q_3} {\bar{H}}_{q_4 q_5 q_6}
\big) \Gamma^{\ell_1 \ell_2}
\nonumber \\
&+& \big({1 \over 12} \Lambda_{\ell_1} Y_{\ell_2 \ell_3 \ell_4}
-{1 \over 4} Y^i{}_{\ell_1 \ell_2} Y_{i \ell_3 \ell_4}
-{1 \over 192} \Phi_{\ell_1} {\bar{H}}_{\ell_2 \ell_3 \ell_4}
-{1 \over 192} {\bar{\Phi}}_{\ell_1} H_{\ell_2 \ell_3 \ell_4}
\nonumber \\
&-&{1 \over 64} H^i{}_{\ell_1 \ell_2} {\bar{H}}_{i \ell_3 \ell_4} \big) \Gamma^{\ell_1 \ell_2 \ell_3 \ell_4}
\bigg) \eta_- \rangle
\nonumber \\
&+& {\rm Re \ } \int_{\cal{S}} \langle \eta_- , \bigg(-{7i \over 16} \Lambda^i \Phi_i - {7 \over 32} h_i \Phi^i
+{11i \over 48} Y_{\ell_1 \ell_2 \ell_3} H^{\ell_1 \ell_2 \ell_3}
\nonumber \\
&+& \big({5i \over 96} \Lambda_{\ell_1} H_{\ell_2 \ell_3 \ell_4}+{5 \over 192} h_{\ell_1} H_{\ell_2 \ell_3 \ell_4}
-{5i \over 48} \Phi_{\ell_1} Y_{\ell_2 \ell_3 \ell_4} -{3i \over 32} Y^i{}_{\ell_1 \ell_2} H_{i \ell_3 \ell_4} \big)
\Gamma^{\ell_1 \ell_2 \ell_3 \ell_4} \bigg) \eta_- \rangle~.
\nonumber \\
\eea

On combining ({\ref{ttr3}}) with ({\ref{termx2c}}) and ({\ref{termx3c}}),
one obtains
\bea
\label{diracdif1}
{\cal{I}} &=&  -{1 \over 2} \int_{\cal{S}} \parallel {\cal A}^{(-)} \eta_-\parallel^2 +
{\rm Re \ } \int_{\cal{S}} \langle \eta_- , \big(-{1 \over 2} h_j \Gamma^j +{i \over 6} Y_{\ell_1 \ell_2 \ell_3}
\Gamma^{\ell_1 \ell_2 \ell_3} \big)  {\cal D}^{(-)}\eta_-\rangle
\cr &&
+ {\rm Re \ } \int_{\cal{S}} \langle \eta_- , \big({3 \over 8} \Phi_j \Gamma^j +{1 \over 48} H_{\ell_1 \ell_2 \ell_3}
\Gamma^{\ell_1 \ell_2 \ell_3} \big) C*{\cal D}^{(-)}\eta_-\rangle~,
\eea

where we have made use of the identity
\bea
\int_{\cal{S}} \tn_i  \langle \eta_- , \Gamma^{ij} \big( - \tn_j \eta_- +{i \over 2} \Lambda_j \eta_- \big) \rangle =0~.
\eea

The expression ({\ref{diracdif1}}) establishes the Lichnerowicz identity.
Suppose then that we impose the horizon Dirac equation
({\ref{dirac1x}}). Then  ({\ref{diracdif1}}) implies that
\bea
 \int_{{\cal{S}}} \parallel \nabla^{(-)}\eta_-\parallel^2
= -{1 \over 2} \int_{\cal{S}}  \parallel {\cal A}^{(-)} \eta_-\parallel^2 ~.
\eea
As the LHS is non-negative, whereas the RHS is non-positive, both sides must vanish.
The vanishing of the LHS implies the horizon KSE ({\ref{skse1x}}), and the vanishing of the
RHS implies ({\ref{alg3x}}).


\appendix{Proof of the preservation of fluxes by Killing vectors}
\label{preservationf}

In this appendix we will give a short proof of the fact that the  vector fields constructed as bi-linears of Killing spinors preserve all fields  of the theory. This is a consequence
 of the results of \cite{iibspingeom}, and a concise proof in the string frame has been given in \cite{figueroa}. Here we shall outline a proof in the Einstein frame for completeness. This relies
  only on the application of the KSEs   (\ref{gkse}) and (\ref{akse}) and thus holds in general.

It is convenient to introduce the following notation
\bea
&& \alpha^{IJ}_{B_1 \cdots B_k} \equiv B(\epsilon^I, \Gamma_{B_1  \cdots B_k}\epsilon^J)~, \notag \\
&& \sigma^{IJ}_{B_1 \cdots B_k} \equiv B(C*\epsilon^I, \Gamma_{B_1  \cdots B_k} C*\epsilon^J)~,  \\
&& \tau^{IJ}_{B_1 \cdots B_k} \equiv B(\epsilon^I, \Gamma_{B_1  \cdots B_k} C*\epsilon^J)~, \notag
\eea
where the inner product $B(\epsilon^I, \epsilon^J) \equiv \langle  \Gamma_0 C* \epsilon^I, \epsilon^J \rangle$ is antisymmetric, i.e.~$B(\epsilon^I, \epsilon^J)=-B(\epsilon^J, \epsilon^I)$ and all $\Gamma$-matrices are anti-Hermitian with respect to this inner product, i.e.~$B(\Gamma_A \epsilon^I,\epsilon^J)=-B(\epsilon^I,\Gamma_A \epsilon^J)$.  See  \cite{iibspingeom} for the conventions.

Denoting $\alpha^{IJ}_{B_1 \cdots B_k} =\alpha^{IJ}_{(k)}$ the bilinears have the symmetry properties
\bea
\alpha^{IJ}_{(k)} = \alpha^{JI}_{(k)}~,\quad \sigma^{IJ}_{(k)} = \sigma^{JI}_{(k)} \quad  k=1,2,5~, \notag \\
\alpha^{IJ}_{(k)} = -\alpha^{JI}_{(k)}~,\quad \sigma^{IJ}_{(k)} = -\sigma^{JI}_{(k)} \quad  k=0,3,4~,
\eea
and their complex conjugates satisfy the following relations
\bea
&&\bar \alpha^{IJ}_{(k)} = \sigma^{IJ}_{(k)}~, \notag\\
&&\bar \tau^{IJ}_{(k)} = - \tau^{JI}_{(k)} \quad k=1,2,5~, \\
&&\bar \tau^{IJ}_{(k)} = \tau^{JI}_{(k)} \quad~~k=0,3,4 \notag ~.
\eea

First we verify that there is a 1-form bi-linear whose associated vector is Killing.  We write the gravitino KSE as
\bea
&&(\nabla_A + \Sigma_A)\epsilon = 0~,
\eea
where
\bea
&& \Sigma_A = -\frac{i}{2} Q_A +\frac{i}{48} F_{A C_1 C_2 C_3 C_4}\Gamma^{C_1 C_2 C_3 C_4} \notag \\
&&\qquad\qquad - \frac{1}{96} (G_{C_1 C_2 C_3}\Gamma_{A}{}^{C_1 C_2 C_3} -9 G_{A C_1 C_2} \Gamma^{C_1 C_2})C* ~,
\eea
which we use to replace covariant derivatives with fluxes and $\Gamma$-matrices. The 1-form bilinear associated with the Killing vector is $\tau^{(IJ)}_A e^A$, which we see by computing
\bea
\nabla_A \tau^{(IJ)}_B &=& \nabla_A B(\epsilon^{(I}, \Gamma_B C*\epsilon^{J)}) \notag \\
&=& B(\nabla_A \epsilon^{(I}, \Gamma_B C*\epsilon^{J)}) +B( \epsilon^{(I}, \Gamma_B C*\nabla_A\epsilon^{J)}) \notag \\
&=& -B(\Sigma_A \epsilon^{(I}, \Gamma_B C*\epsilon^{J)}) -B( \epsilon^{(I}, \Gamma_B C*\Sigma_A\epsilon^{J)}) \notag \\
&=& B(\Gamma_B C*\epsilon^{(I},\Sigma_A \epsilon^{J)}) -B( \epsilon^{(I}, \Gamma_B C*\Sigma_A\epsilon^{J)})  \\
&=& -\bar B(\epsilon^{(I},C* \Gamma_B \Sigma_A \epsilon^{J)}) -B( \epsilon^{(I}, \Gamma_B C*\Sigma_A\epsilon^{J)}) \notag \\
&=& -2  {\rm Re \ } B(\epsilon^{(I}, C* \Gamma_{B}\Sigma_A \epsilon^{J)})\notag \\
&=& {\rm Re }\Big(-\frac{3}{8} G_{A B}{}^C \bar\alpha^{IJ}_C - \frac{1}{48}G^{C_1 C_2 C_3}\bar \alpha^{IJ}_{A B C_1 C_2 C_3} + \frac{i}{6}F_{A B}{}^{C_1 C_2 C_3} \tau^{IJ}_{C_1 C_2 C_3}\Big) \notag ~.
\eea
Since the resulting expression is antisymmetric in its free indices we find that $\nabla_{(A} \tau^{(IJ)}_{B)}=0 $ and hence the vector associated with $\tau^{(IJ)}_A e^A$ is Killing.

Note that the dilatino KSE (\ref{akse})
\bea
{\cal A} \epsilon \equiv \Big(P_A \Gamma^A (C*) +\frac{1}{24} G_{A_1 A_2 A_3}\Gamma^{A_1 A_2 A_3}\Big)\epsilon = 0~,
\eea
implies that
\bea
0  = B(\epsilon^{(I},{\cal A} \epsilon^{J)}) = P^A \tau^{(IJ)}_A ~,
\eea
and hence $i_K P=0$, where $K=\tau^{(IJ)}_A e^A$ denotes the 1-form associated with the Killing vector. With this relation it follows that the Killing vector leaves $P$ and $dQ$ invariant up to a $U(1)$ transformation:
\bea
{\cal L}_K P = i_K dP + d (i_K P) =  2 i ~i_K (Q \wedge P) = 2i (i_K Q) P ~,
\eea
and
\bea
{\cal L}_K dQ = d i_k dQ = -i d i_K (P \wedge \bar P) = 0~,
\eea
where we have used the Bianchi identities for $P$ and $Q$
\bea
&&dP = 2 i Q\wedge P~,~~~dQ = -i P \wedge \bar P ~.
\eea

To see that the 3-form flux $G$ is preserved we need to analyse the 1-form bi-linear which is not related to the Killing vector, i.e. $\alpha^{IJ}_A e^A$. As above, we find that\footnote{The terms containing the 5-form flux $F$ and a 5-form bi-linear vanish due to the self-duality of $F$.}
\bea
\nabla_{[A} \alpha^{IJ}_{B]} &=& -2 B(\epsilon^{(I},\Gamma_B \Sigma_A \epsilon^{J)}) \notag \\
&& = - \frac{1}{2} G_{AB}{}^C \tau^{(IJ)}_C + i Q_{[A} \alpha^{IJ}_{B]} - P_{[A} \bar\alpha^{IJ}_{B]}~,
\label{3formeq}
\eea
or equivalently
\bea
d\alpha = -i_K G + i Q\wedge \alpha - P \wedge \bar\alpha ~,
\label{3formeq2}
\eea
where we have suppressed the indices labelling the Killing spinors on $\alpha$. To arrive at (\ref{3formeq}) we have used the dilatino KSE in the form
\bea
B(C* \epsilon^I, \Gamma_{AB} {\cal A} \epsilon^J)~,
\eea
to cancel bi-linears other than $\alpha$ and the 1-form bi-linear associated with the Killing vector. By taking the exterior derivative of (\ref{3formeq2}), and resubstituting the expression for $d\alpha$, it follows that
\bea
{\cal L}_K G = i (i_K Q) G ~,
\eea
where we have used the Bianchi identity for $G$
\bea
dG = i Q\wedge G - P \wedge \bar G ~.
\label{GBI}
\eea

In now remains to verify that also the 5-form flux $F$ is preserved. Since we have already analyzed the possible 1-form bi-linears, we proceed to the 3-form bi-linears. There is only one 3-form bi-lilnear which is symmetric under the exchange of spinors, namely $\tau^{(IJ)}_{B_1 B_2 B_3} e^{B_1}\wedge e^{B_2}\wedge e^{B_3}$. As above, we find
\bea
d\tau = -4 i_K F +\frac{i}{2} G\wedge \bar \alpha -\frac{i}{2} \bar G \wedge \alpha ~.
\eea
Taking the exterior derivative of this expression yields
\bea
{\cal L}_K F = 0 ~,
\eea
where we have used the expression above for $d\alpha$ and the Bianchi identity for $G$ and $F$
\bea
dF = \frac{i}{8}G\wedge \bar G ~.
\eea
For the computations in this appendix the Mathematica package GAMMA \cite{Gran:2001yh} has been used.

\end{document}